\documentclass[12pt,fleqn,a4paper]{article}
\usepackage{epsfig,amssymb,amsthm,amsmath,mathrsfs}

\newcommand{\rem}[1]{} 

\newcommand{\R}{{\mathbb{R}}}

\newcommand{\Z}{{\mathbb{Z}}}

\newcommand{\res}{\mathop{\mathrm{Res}}}

\newcommand{\eps}{\varepsilon}
\newcommand{\imag}{\mathrm{i}}

\newcommand{\id}{{\rm id}}
\newcommand{\dee}{{\rm d}}

\newtheorem{theorem}{Theorem}[section]
\newtheorem{lemma}[theorem]{Lemma}
\newtheorem{proposition}[theorem]{Proposition}
\newtheorem{corollary}[theorem]{Corollary}
\newtheorem{definition}[theorem]{Definition}

\begin{document}

\title{Geodesic flow on three dimensional ellipsoids with equal semi-axes}

\author
{
{\protect\normalsize Chris M. Davison, Holger R.\ Dullin} \\
{\protect\footnotesize\protect\it
Department of Mathematical Sciences, Loughborough University} \\[-2mm]
{\protect\footnotesize\protect\it
Leicestershire, LE11 3TU, United Kingdom\footnote{ email:
c.m.davison@lboro.ac.uk, h.r.dullin@lboro.ac.uk
}.}
}

\date{\protect\normalsize 2 November 2006}

\maketitle

\begin{abstract}
Following on from our previous study of the geodesic flow on three dimensional ellipsoid with equal middle semi-axes, 
here we study the remaining cases: Ellipsoids with two sets of equal semi-axes with $SO(2) \times SO(2)$ symmetry, 
ellipsoids with equal larger or smaller semi-axes with $SO(2)$ symmetry, and ellipsoids with three  semi-axes 
coinciding with $SO(3)$ symmetry. 
All of these cases are Liouville-integrable, and reduction of the symmetry leads to singular reduced 
systems on lower-dimensional ellipsoids. The critical values of the energy-momentum maps and their singular
fibers are completely classified. In the cases with $SO(2)$ symmetry there are corank 1 degenerate critical 
points; all other critical points are non-degenreate. We show that in the case with $SO(2) \times SO(2)$ symmetry 
three global action variables exist and the image of the energy surface under the energy-momentum map is a 
convex polyhedron. The case with $SO(3)$ symmetry is non-commutatively integrable, and we show that the 
fibers over regular points of the energy-casimir map are $T^2$ bundles over $S^2$.
\end{abstract}

\section{Introduction}

The geodesic flow on the ellipsoid is the classical example of a non-trivial separable and thus Liouville 
integrable Hamiltonian system. It is the prime example in Jacobi's famous 
``Vorlesungen \"uber Dynamik'' \cite{jacobi84} and may be considered as his motivation to develop 
Hamilton-Jacobi theory and the solution of the Abel-Jacobi inversion problem.
Its modern treatment was pioneered by the Z\"urich school, 
namely by Moser \cite{moser80} and Kn\"orrer \cite{knorrer80,knorrer82}, generalising to the $n$-ellipsoid and providing
smooth integrals and the general solution in terms of $\theta$-functions
for the generic case of an $n$-ellipsoid with pair-wise distinct semi-axes.

Zung \cite{zung96} carried out an excellent general approach to the topology of St\"ackel systems, including the geodesic flow on the ellipsoid with distinct semi-axes.  In \cite{BDD} we extended his results to the degenerate case.  We reviewed the geodesic flow on the three dimensional ellipsoid with distinct semi-axes, and analysed what happens in the degenerate case where we set the two middle semi-axes equal.  We found that the topology of the critical values in the image of the energy-momentum map changes and the torus-bundle over the regular values is non-trivial and has monodromy.  This showed that by making the system simpler (i.e. more symmetric) it becomes more complicated (i.e. have a non-trivial torus bundle).

In this study we complete the classification of the geodesic flow on the remaining three dimensional ellipsoids with equal semi-axes, namely the case with two sets of equal semi-axes and $SO(2) \times SO(2)$ symmetry, 
the case with either largest or smallest equal semi-axes and $SO(2)$ symmetry, 
and the case with three equal semi-axes and $SO(3)$ symmetry.
In section 2 we give a brief review of the geodesic flow on general ellipsoids in order to fix our notation.
The case with $SO(2) \times SO(2)$ symmetry is described in section 3.
Symmetry reduction leads to a (singular) system on the ellipse with an effective potential.
The symmetry is a torus action which gives rise to a global momentum map.
The image of the energy surface under this momentum map is a convex polyhedron. 
Equivalently the image of the energy surface under the energy-momentum map is that
convex polyhedron, except that now the generic fibers are three-tori (by Liouville-Arnold).
The convexity of the image is related to results of Atiyah \cite{atiyah82} and Guillemin-Sternberg \cite{guillemin82} 
on the convexity of the image of the momentum map of torus actions. However, their results apply to 
compact symplectic manifolds only. Generalizations to the non-compact case exist \cite{sjamaar98,lerman03}, but 
our example seems to be new. In the end of section 3 we prove the existence of three smooth global action 
variables for this situation. The fact that the third action is smooth (and does not have monodromy) is 
not obvious, since the natrual action defined via the cycles from the separation of variables is only
continuous, but not smooth.
  
The ellipsoid with the $SO(2)$ symmetry and equal smallest semi-axies is described in section 4.
The system reduces to a (singular) system on the 2-ellipsoid with additional potential 
and a billiard wall inserted in a plane {\em not} containing the umbilic points.
Two singular values have critical points of corank 1 that are degenerate, while all other singular 
values are non-degenerate. The topology of the fibres is deduced using Poincar\'e sections.
Section 5 gives similar results for the case with equal largest semi-axes.

In the final section the two cases with three equal semi-axes are described using singular reduction 
to one degree of freedom. In addition to studying the energy-momentum map in this case another
interesting object is the energy-casimir map, since the system is non-commutatively 
integrable (or superintegrable). 
We show that the fibers of the generic point in the image of the energy-casimir map
is a 2-torus bundle over the 2-sphere. 

\section{Review of the the geodesic flow on generic 3-ellipsoids}

The geodesic flow on the generic 3-ellipsoid with distinct semi-axes was described in more detail in \cite{BDD}.  
Here we will briefly review some of the terminology to provide a foundation for the examples described here.
The standard form of the 3-ellipsoid embedded in $\R^4$ with coordinates $x = (x_0, x_1, x_2, x_3)$ and semi-axes $\sqrt{\alpha_i}$, for $0 < \alpha_0 \le \alpha_1\le \alpha_2 \le \alpha_3 $, is
\[
C_1 = 
    \frac{x_0^2}{\alpha_0} + 
    \frac{x_1^2}{\alpha_1} + 
    \frac{x_2^2}{\alpha_2} + 
    \frac{x_3^2}{\alpha_3}  
    - 1 = 0 \,.
\]
For the generic non-degenerate ellipsoid the semi-axes are distinct.
The Lagrangian of a free particle with mass 1 is 
$L = \frac12( \dot x_0^2 +  \dot x_1^2 +  \dot x_2^2 + \dot x_3^2 )$.  A Hamiltonian description can be obtained by introducing momenta $y_i = \dot x_i$ and enforcing the constraint by replacing the standard symplectic structure $\dee x \wedge \dee y$ by a Dirac bracket. 
The Dirac bracket preserves the ellipsoid $C_1 = 0$ and its tangent space
\[
C_2 = 
\frac{x_0 y_0}{\alpha_0} +
\frac{x_1 y_1}{\alpha_1} +
\frac{x_2 y_2}{\alpha_2} +
\frac{x_3 y_3}{\alpha_3}  = 0 \,.
\]
We define also the following notation
\[
    D =
    \frac{x_0^2}{\alpha_0^2} + 
    \frac{x_1^2}{\alpha_1^2} + 
    \frac{x_2^2}{\alpha_2^2} + 
    \frac{x_3^2}{\alpha_3^2}  
    = \frac12 \sum \frac{\partial C_1}{\partial x_i} \frac{ \partial C_2}{\partial y_i}.
\]
We can generalise the constraints $C_1$ and $C_2$, and the factor $D$, for a $n-1$ ellipsoid ${\cal E}$ embedded in $\R^n$. Lifting to the cotangent bundle, we have coordinates $x=(x_0,\ldots,x_{n-1})$ and conjugate momenta $y=(y_0,\ldots,y_{n-1})$ for $T^*{\cal E}$ embedded in $T^*\R^n$. For this generic case the Dirac bracket with Casimirs $C_1$ and $C_2$ is given by 
\begin{equation}
\label{eqn:dirac}
\left\{x_i,x_k\right\}_{2n} = 0, \qquad 
\left\{x_i,y_k\right\}_{2n} = \delta_{ik} - \frac {x_ix_k} {D \alpha_i \alpha_k }, \qquad 
\left\{y_i,y_k\right\}_{2n} = - \frac { x_i y_k - x_k y_i} {D \alpha_i \alpha_k} \,,
\end{equation}
where the subscript $2n$ indicates the dimension of the embedding space $T^*\R^n$. 

Returning to the three dimensional ellipsoid, the Hamiltonian is $H = \frac12( y_0^2 + y_1^2 + y_2^2 + y_3^2)$
and the equations of motion are
\begin{equation}
    \dot x_i = \{x_i,H\}_8, \qquad \dot y_i = \{y_i,H\}_8, \quad i = 0,1,2,3.
\end{equation}
The vector field generated by $H$ is denoted by $X_H$.

The system is Liouville integrable with smooth global integrals 
(in the generic case of distinct semi-axes) first found by Uhlenbeck and Moser \cite{moser80}
\begin{equation}
\label{eqn:gen3dints}
F_i = y_i^2 + \sum_{k=1, k \ne i}^{n} \frac {\left(x_iy_k - x_ky_i\right)^2} {\alpha_i - \alpha_k}, 
\quad i =  0, \dots, 3 \,.
\end{equation}
On the symplectic leaf of the Dirac bracket given by $C_1= C_2 = 0$
they are related by $\sum F_i/\alpha_i = 0$ and they have pair-wise vanishing bracket
\cite{moser80}.
The integrals $F_i$ are related to the Hamiltonian by $H = \frac12( F_0 + F_1 + F_2 + F_3)$.

\begin{figure}
\centerline{\includegraphics[width=14cm]{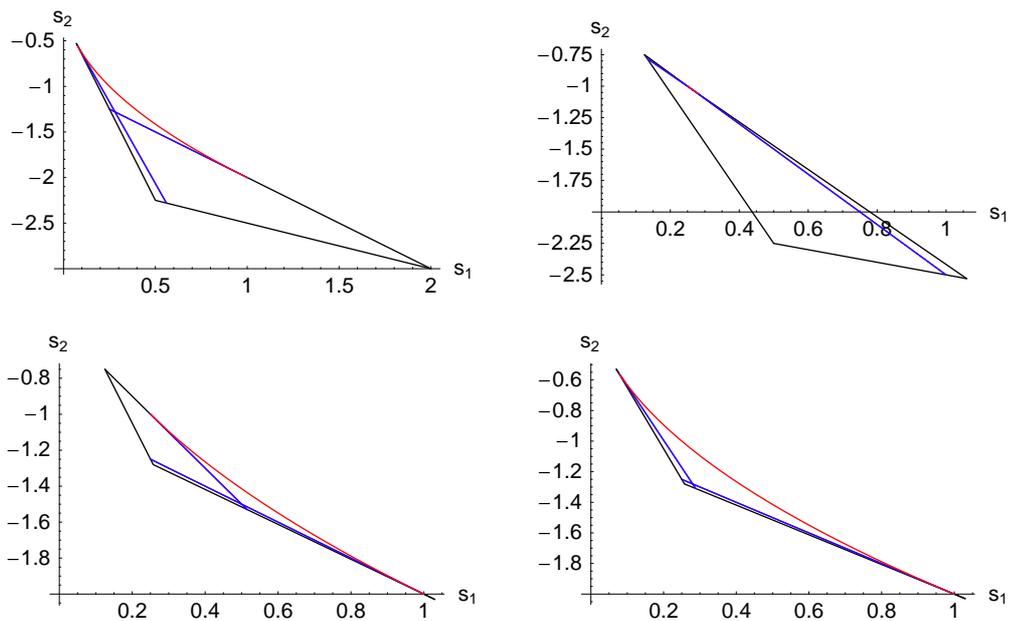}}
\caption{Bifurcation Diagram of almost degenerate Ellipsoids with 
$\alpha_i = (1/4, 1/4 + \eps, 1, 2)$, 
$(1/4, 1/2, 1/2 + \eps, 2)$, 
$(1/4, 1/2, 1, 1 + \eps)$, 
$(1/4, 1/4+\eps, 1, 1 + \eps)$, where $\eps = 0.03$.
} 
\label{DegenBifDiags}
\end{figure}
\noindent
As a preparation for the cases studied here, consider the bifurcation diagram at constant energy for the generic case for the four cases in which the semi-axes nearly coincide, as shown in figure \ref{DegenBifDiags}.  The bifurcation diagrams are constructed by separating the variables using an ellipsoidal coordinate system and Hamilton-Jacobi theory, as described in \cite{BDD}.  From top left to bottom right the cases are close to equal smallest axes, equal middle axes, equal largest axes, and equal smallest and largest axes, also denoted by 211, 121, 112, and 22.  For the 22 case we see that the image of the momentum map has only a single chamber.  For the 112 and 211 cases two chambers are present in the diagrams.  We must therefore analyse these cases in further detail.

\section{Geodesic flow for the ellipsoid with $SO(2) \times SO(2)$ symmetry}

The three-ellipsoid ${\cal E}$ with two pairs of equal axes is defined by
\begin{equation}
\frac {x_0^2} {\alpha_1} + \frac {x_1^2} {\alpha_1} + \frac {x_2^2} {\alpha_2} + \frac {x_3^2} {\alpha_2} = 1,
\end{equation}
where $0 < \alpha_1 < \alpha_2$.  The $SO(2) \times SO(2)$ symmetry group action is given by
\begin{equation}
\Psi(x, y; \theta_1, \theta_2) = (\tilde x, \tilde y),
\end{equation}
where
\begin{align*}
\tilde x & = (x_0\cos \theta_1-x_1\sin \theta_1, x_0\sin \theta_1+x_1\cos \theta_1, x_2\cos \theta_2-x_3\sin \theta_2, x_2\sin \theta_2+x_3\cos \theta_2), \\
\tilde y & = (y_0\cos \theta_1-y_1\sin \theta_1, y_0\sin \theta_1+y_1\cos \theta_1, y_2\cos \theta_2-y_3\sin \theta_2, y_2\sin \theta_2+y_3\cos \theta_2).
\end{align*}
$\Psi$ is generated by the momentum map ${\cal M} = (J_1, J_2) : M \rightarrow \mathbb{R}^2$, where $J_1 = x_0y_1 - x_1y_0$ and $J_2 = x_2y_3 - x_3y_2$ are the angular momenta.

The Dirac bracket is given by \eqref{eqn:dirac} with $\alpha_0=\alpha_1$ and $\alpha_3=\alpha_2$.

\subsection{Liouville Integrability}
\begin{theorem} {\em } \label{thm:ALI2}
The geodesic flow on the ellipsoid with two pairs of equal axes is Liouville integrable.  
Constants of motion are the energy $H=\frac 12 (y_0^2+y_1^2+y_2^2+y_3^2)$, and angular momenta $J_1$ and $J_2$.
\end{theorem}
\begin{proof}
$H$, $J_1$ and $J_2$ commute with respect to the Dirac bracket \eqref{eqn:dirac}.  Moreover they are independent almost everywhere since they are polynomials and independent at, e.g.,  
$x=(\sqrt{\alpha_1}, 0, 0, 0)$, $y=(0, 0, 1, 0)$.
\end{proof}

We note that we also have smooth globally defined integrals $G_1$ and $G_2$.  $G_1$ is defined by $G_1=F_0+F_1$, where the $F_i$ are as in \eqref{eqn:gen3dints} for the generic three dimensional ellipsoid, and letting $\alpha_0 \rightarrow \alpha_1$.  Similarly $G_2$ is defined by $G_2=F_2+F_3$ for $\alpha_3 \rightarrow \alpha_2$.  Hence
\begin{equation}
G_1 = y_0^2+y_1^2 + \frac {1} {\alpha_1-\alpha_2}\left[(x_0y_2-x_2y_0)^2 + (x_0y_3-x_3y_0)^2 + (x_1y_2-x_2y_1)^2 + (x_2y_3-x_3y_2)^2 \right],
\nonumber
\end{equation}
\begin{equation}
G_2 = y_2^2+y_3^2 + \frac {1} {\alpha_2-\alpha_1}\left[(x_0y_2-x_2y_0)^2 + (x_0y_3-x_3y_0)^2 + (x_1y_2-x_2y_1)^2 + (x_2y_3-x_3y_2)^2 \right].
\nonumber
\end{equation}
We have the relations $2H=G_1+G_2$ and
\begin{equation}
\frac {G_1} {\alpha_1} + \frac {G_2} {\alpha_2} - \frac {J_1^2} {\alpha_1^2} - \frac {J_2^2} {\alpha_2^2} =0.
\nonumber
\end{equation}
These integrals will be used later in the proof of the non-degeneracy of the singular points of the momentum map.

The group action $\Psi$ has the invariants
\begin{equation}
\pi_1^1 = x_0^2 + x_1^2, \qquad \pi_2^1 = y_0^2 + y_1^2, \qquad \pi_3^1 = x_0y_0 + x_1y_1, \qquad \pi_4^1 = x_0y_1 - x_1y_0,
\end{equation}
\begin{equation}
\pi_1^2 = x_2^2 + x_3^2, \qquad \pi_2^2 = y_2^2 + y_3^2, \qquad \pi_3^2 = x_2y_2 + x_3y_3, \qquad \pi_4^2 = x_2y_3 - x_3y_2,
\end{equation}
related by
\begin{equation}
\pi_1^1\pi_2^1 - (\pi_3^1)^2 - (\pi_4^1)^2 = 0, \qquad \pi_1^2\pi_2^2 - (\pi_3^2)^2 - (\pi_4^2)^2 = 0.
\label{eqn:reln22}
\end{equation}
$\Psi$ is not free, but has fixed points at the origin in the 
$x_0, x_1, y_0, y_1$ plane for $\theta_2 = 0$ and in the $x_2, x_3, y_2, y_3$ plane for $\theta_1 = 0$,
or both. These are the only fixed points, and they occur for zero angular momentum only.
When $J_1=\pi_4^1=j_1 \ne 0$ and $J_2 = \pi_4^2 = j_2 \ne 0$ the fixed points are not in ${\cal M}^{-1}(j_1, j_2)$, so excluding $j_1 j_2 = 0$ we can reduce by $\Psi$ and obtain a regular reduced system.

\begin{theorem} \label{thm:regredn2}
A set of reduced coordinates $(\xi_1, \xi_2, \eta_1, \eta_2)$ is defined on the reduced phase space ${\cal M}^{-1}(j_1, j_2) / \Psi$, for $j_1 j_2 \ne 0$, by the formulae
\[
   \xi_1=\sqrt{\pi_1^1}, \qquad \xi_2=\sqrt{\pi_1^2}, \qquad
   \eta_1 =  \frac{\pi_3^1}{\sqrt{\pi_1^1}}, \qquad \eta_2 =  \frac{\pi_3^2}{\sqrt{\pi_1^2}}.
\]
The reduced coordinates satisfy the Dirac bracket in $\R^4[\xi,\eta]$, i.e. $\{.,.\}_4$ as defined in \eqref{eqn:dirac}.
\noindent
The mapping $R:\R^8[x,y] \to \R^4[\xi,\eta]$ is Poisson from $\R^8$ with $\{.,.\}_8$ to $\R^4$ with $\{.,.\}_4$ and the reduced system has reduced Hamiltonian
\begin{equation} \label{eqn:HamRed}
\hat H = \frac {1} {2} \left(\eta_1^2 + \eta_2^2\right) + \frac{1} {2} \left(\frac {j_1^2} {\xi_1^2} + \frac {j_2^2} {\xi_2^2}\right).
\end{equation}
\end{theorem}

\begin{proof}
Define a set of coordinate on $\R^4[\xi,\eta]$ as shown. The new variables $(\xi_1, \xi_2, \eta_1, \eta_2)$ are invariant under the group action $\Psi$. 
The Poisson property of the map $R$, i.e. $\{ f \circ R, g \circ R\}_8 = \{f, g\}_4 \circ R$
follows from direct computation of the basic brackets, e.g.
$\{ \xi_1, \xi_2 \}_4 = \{ \sqrt{x_0^2+x_1^2}, \sqrt{x_2^2+x_3^2} \}_8 =0$, etc.
Using the identities
\begin{equation}
\pi_2^1 = \frac {J_1^2} {\xi_1^2} + \eta_1^2, \qquad \pi_2^2 = \frac {J_2^2} {\xi_2^2} + \eta_2^2,
\end{equation}
the new Hamiltonian is found, which is that of the geodesic flow on the ellipse plus an effective potential.
In the new variables the two Casimirs are
\begin{equation}
\frac {\xi_1^2} {\alpha_1} + \frac {\xi_2^2} {\alpha_2} = 1, \qquad \frac {\xi_1 \eta_1} {\alpha_1} + \frac {\xi_2 \eta_2} {\alpha_2} = 0 \,.
\label{eqn:newcasimir}
\end{equation}
\end{proof}

\begin{figure}
\centerline{\includegraphics[width=6cm]{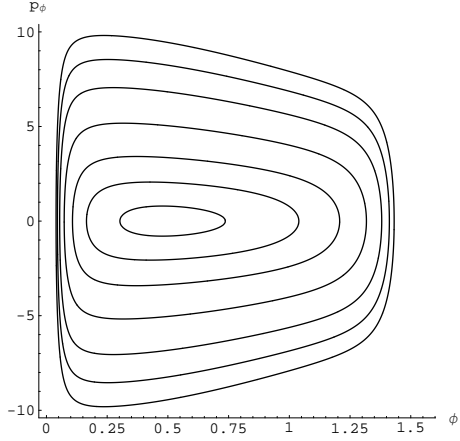}}
\caption{Contour plot of reduced Hamiltonian with $\alpha_1=1,\alpha_2=2, j_1=1, j_2=0.4$}
\label{fig:Action22}
\end{figure}

The condition $j_1 j_2 \not = 0$ insures that the transformation to $\eta_i$ and the 
Hamiltonian are smooth. 
The reduced Casimirs show that the reduced phase space is $T^* S^1$.
Symplectic coordinates $(\phi, p_\phi)$ may be chosen as
\begin{equation}
\begin{aligned}
\xi_1 &= \sqrt{\alpha_1}\cos(\phi), &
 \xi_2 & = \sqrt{\alpha_2}\sin(\phi), \\
\eta_1 &= \frac {-\sqrt{\alpha_1}\sin(\phi)p_\phi} {\alpha_1\sin^2(\phi)+\alpha_2\cos^2(\phi)},& 
\eta_2 &= \frac {\sqrt{\alpha_2}\cos(\phi)p_\phi} {\alpha_1\sin^2(\phi)+\alpha_2\cos^2(\phi)}.
\end{aligned}
\end{equation}
These are symplectic coordiantes as  $\{\phi, p_\phi \}=1$.  Expressing the reduced Hamiltonian in these coordinates gives
\begin{equation} \label{SympHam}
\hat H = \frac {1} {2} \left(\frac {p_\phi^2} {\alpha_1\sin^2(\phi)+\alpha_2\cos^2(\phi)} + \frac {j_1^2} {\alpha_1\cos^2(\phi)} + \frac {j_2^2} {\alpha_2\sin^2(\phi)} \right).
\end{equation}
The contour plot for the reduced Hamiltonian is shown in figure \ref{fig:Action22}.

Notice that $\xi_i > 0$, so that only one quarter of the circle, $0 < \phi < \pi/2$  
is the reduced configuration space. When allowing the full circle the quotient
by the $\Z_2 \times \Z_2$ action that flips the signs of $\xi_i, \eta_i$, $i = 1, 2$
gives back the quarter circle. Here this action is trivial, but for the singular 
reduction it will become more interesting.

\begin{figure}
\centerline{\includegraphics[width=6cm]{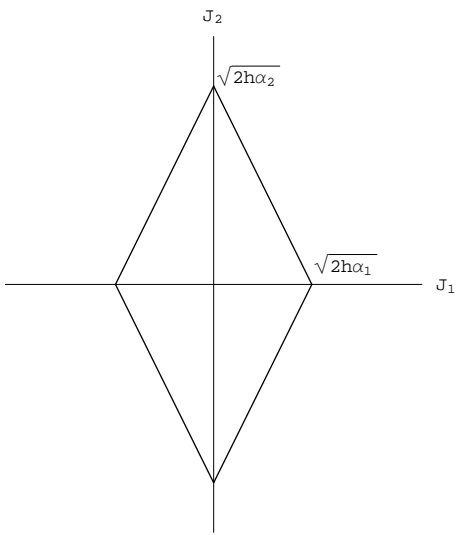}}
\caption{Bifurcation diagram for the ellipsoid with $\alpha_0 = \alpha_1 < \alpha_2 = \alpha_3$ for $\alpha_1=1, \alpha_2=2, h=1$.}
\label{BifDiag22}
\end{figure}

The reduced Hamiltonian has one degree of freedom, and equilibrium points of 
the reduced system in general correspond to two-tori in full phase space. 
Since the reduced Hamiltonian grows monotonously with the momenta, for fixed 
energy these equilibrium points occur at minima of $\hat H$.
Conversely, the image of the energy surface under the momentum map has as a boundary
the values of momenta for which equilibria are found in the reduced system.
\begin{theorem} \label{thm:EM2}
The image of the energy surface ${\cal E}_h=\{(\mathbf{x},\mathbf{y}) \in T^*{\cal E} | H=h\}$ under the momentum map is the convex region bounded by the polygon defined by the lines
\begin{equation}
\pm \frac {j_1} {\sqrt{\alpha_1}} \pm \frac {j_2} {\sqrt{\alpha_2}} = \sqrt{2h}.
\label{eqn:convexpolygon}
\end{equation}
\end{theorem}
\begin{proof}
The boundary of the image of the energy surface under the momentum map corresponds to the equilibrium points of the reduced flow.  Equating $\{\xi_i,\hat H\}$ to zero gives
\begin{equation}
\eta_i - \frac {\xi_i} {\Delta\alpha_1}\left(\frac {\xi_1\eta_1} {\alpha_1} + \frac {\xi_2\eta_2} {\alpha_2}\right) = 0,
\end{equation}
so that $\eta_i=0$, using  the Casimir \eqref{eqn:newcasimir}.  Using this, the equation $\{ \eta_i, \hat H\} = 0$ 
reduces to 
\begin{equation}
\alpha_ij_i^2\left(\frac {\xi_1^2} {\alpha_1^2} + \frac {\xi_2^2} {\alpha_2^2}\right) = \xi_i^4 \left(\frac {j_1^2} {\alpha_1\xi_1^2} + \frac {j_2^2} {\alpha_2\xi_2^2}\right) \,.
\end{equation}
Dividing this equation by $\alpha_i j_i^2 \xi_i^2$ gives $\xi_1^4/\xi_2^4 = \alpha_1 j_1^2 / (\alpha_2 j_2^2)$. Inserting the ansatz $\xi_i^2 = \pm s \sqrt{ \alpha_i } j_i$ with a scaling factor $s$ 
into the Casimir of the ellipse determines $s$ and thus gives the result
\begin{equation}
\xi_i^4 = \frac {\alpha_ij_i^2} {\left(\pm \frac {j_1} {\sqrt{\alpha_1}} \pm \frac {j_2} {\sqrt{\alpha_2}}\right)^2}.
\end{equation}
The sign in the ansatz is chosen so that $\pm j_i$ is positive.
Evaluating the Hamiltonian at these critical points gives the critical values which are the boundaries of the polygon shown in figure \ref{BifDiag22}.
\end{proof}

This result is a generalization of a general result by 
Atiyah \cite{atiyah82} and Guillemin-Sternberg \cite{guillemin82} on the convexity of the image of the 
momentum map. The classical result applies to compact symplectic manifolds 
only, but extensions to certain non-compact situations exist \cite{sjamaar98,lerman03}.
The fibres over the momentum map in the compact case follow from the general theory.
In the non-compact case treated here the fibres are computed explicitly using singular reduction.
The singular reduction also works when $j_1 j_2 = 0$. However, the reduced phase space
is not a smooth manifold in this case. Previously we have seen that in the regular case 
$j_1 j_2 \not = 0$ the reduced phase space is the cotangent bundle over an open 
interval in $\phi$, hence it is diffeomorphic to $\R^2$.

The bifurcation diagram figure~\ref{BifDiag22} can be considered as taking the square root twice of the
diagram in figure~\ref{DegenBifDiags} bottom right. The line $J_1=0$ is where
the lines $F_0 = 0$ and $F_1 = 0$ collapse, and $J_2=0$ is where $F_2=0$ and $F_3=0$ collapse. In the limit coming from the generic case the whole of each line would appear to be critical,
since they are both on the boundary of the image of the energy-momentum map. Recall that the limit of $F_0(\alpha_0 - \alpha_1)$ as $\alpha_1 \rightarrow \alpha_0$ (or $F_1(\alpha_1 - \alpha_0)$)
equals $J_1^2$, but not $J_1$. Similarly $J_2^2$ is the limit of $F_2(\alpha_2-\alpha_3)$. Obviously $J_1^2$ is singular when $J_1=0$, but $J_1$ itself is not, and similarly for $J_2^2$. Thus taking the square root twice of figure~\ref{DegenBifDiags} bottom right gives figure~\ref{BifDiag22}. Upon this transition the critical points along the lines $F_0 = F_1 = 0$ and $F_2=F_3=0$ disappear. The boundary of the convex polyhedron the result of taking the square root twice of the quadratic curve in figure~\ref{DegenBifDiags} bottom right arising from the double roots. The multiplicity of the regular $T^3$
changes from 4 to 1 for every regular point in the image.

\subsection{Singular Reduction}
\begin{theorem} \label{thm:SR2}
The singular reduced space has one or two conical singularities. 
The only singular values of the energy-momentum map for constant energy $H=h$ are the boundaries of the convex polygon 
forming the image of the energy surface under the momentum map.
\end{theorem}
\begin{proof}
We use invariant theory to perform singular reduction in order to find the reduced phase space. The Casimirs for the system expressed in terms of the invariants are
\begin{equation}
C_1 = \frac {\pi_1^1} {\alpha_1} + \frac {\pi_1^2} {\alpha_2} - 1 = 0, \qquad C_2 = \frac {\pi_3^1} {\alpha_1} + \frac {\pi_3^2} {\alpha_2} = 0,
\end{equation}
and expressing the Hamiltonian in terms of the invariants gives
\begin{equation}
H = \frac {1} {2} \left(\pi_2^1 + \pi_2^2\right).
\label{eqn:enginvt22}
\end{equation}
To find the reduced phase space ${\cal M}^{-1}(j_1, j_2) / \Psi$ we eliminate the invariants from the second relation \eqref{eqn:reln22}.  The reduced phase space is then defined by the equations
\begin{equation}
\pi_1^1\pi_2^1 - (\pi_3^1)^2 = j_1^2, \qquad \alpha_2\left(1 - \frac {\pi_1^1} {\alpha_1}\right)\pi_2^2 - \frac {\alpha_2^2} {\alpha_1^2} (\pi_3^1)^2 = j_2^2
\label{eqn:invt22}
\end{equation}
together with the inequalities
\begin{equation}
\pi_1^1 \ge 0, \qquad \pi_2^1 \ge 0, \qquad \pi_2^2 \ge 0.
\end{equation}
Differentiating the five equations with respect to the six variables and computing the 
rank we find that the surfaces defined by these equations are smooth manifolds unless 
one or both of the $j_i$ are zero. When $j_i = 0$ then the non-negativity of $\pi_1^i$ and $\pi_2^i$
implies that all $\pi^i_k = 0$, $k= 1,2, 3$.

We consider the three cases which may occur.   
For $j_1 j_2 \ne 0$ we have the strict inequalities $\pi_1^1 > 0, \pi_2^1 > 0, \pi_2^2 > 0$.  
Thus we can eliminate $\pi_1^1$ and the resulting equation is that of a two-sheeted hyperboloid. 
Only one sheet is relevant due to the positivity of the invariants. This is the $\R^2$ found before
using symplectic coordinates $(\phi, p_\phi)$.

For the singular cases it is best to consider a covering of the reduced phase space.
In \cite{BDD}
this was done for the action generated by a single momentum.
Here we repeat the construction with both momenta. 
The result is that we describe the reduced phase space as that of the 
geodesic flow on the ellipse with coordinates $(\xi, \eta)$ 
quotient by the discrete group $\Z_2 \times \Z_2$.
The group action of the $i$th factor is given by flipping the signs of $\xi_i, \eta_i$.
When $\xi_i=\eta_i = 0$ is not in the reduced phase space this action
merely identifies the four quarters of the circle. 
In the singular case, however, conical singularities appear whenever $j_i = 0$ 
and hence $\xi_i=\eta_i = 0$ is possible.
When only one angular momentum is zero this gives one conical singularity
in the reduced phase space, when both are zero it gives two.
Thus the reduced phase space for $j_1 = j_2 = 0$ is the canoe \cite{cushman97}.

We consider the energy surface in the reduced phase space.  Fixing the energy $H=h$ in the Hamiltonian \eqref{eqn:enginvt22}, we may use it to eliminate a further invariant in \eqref{eqn:invt22} because the Hamiltonian is linear in the invariants.  
Changing $h$ leaves the energy surface invariant up to a scaling of momenta, which 
is always the case for geodesic flows.
Fixing one positive $h$ gives the equations
\begin{equation}
Q_1 = \pi_1^1\pi_2^1 - (\pi_3^1)^2 - j_1^2 = 0, \qquad Q_2 = \alpha_2\left(1 - \frac {\pi_1^1} {\alpha_1}\right)\left(2h - \pi_2^1\right) - \frac {\alpha_2^2} {\alpha_1^2} (\pi_3^1)^2 - j_2^2 = 0 \,.
\end{equation}

The singular points are where $\nabla Q_1 \times \nabla Q_2 = 0$. The only real solution to this equation is
\begin{equation}
\pi_3^1 = 0, \qquad \pi_2^1 = \frac {2h\pi_1^1} {\alpha_1}.
\label{eqn:soln1}
\end{equation}
Finding the remaining invariants from the Casimirs and energy equation 
and substituting into the relations gives
\begin{equation}
j_1^2 = \frac {2h(\pi_1^1)^2} {\alpha_1}, \qquad j_2^2 = 2h\alpha_2\left(1 - \frac {\pi_1^1} {\alpha_1}\right)^2.
\label{eqn:soln2}
\end{equation}
Eliminating $\pi_1^1$ from these equations gives us the equation \eqref{eqn:convexpolygon} for the convex polygon found earlier.  In other words the only singular values are on the boundary of the polygon.  The important thing to note here is that the coordinates axes in the image of the 
momentum map are not singular values, unless at the corners of the polygon.
In particular also the origin $j_1 = j_2 = 0$ is not a singular value. 
The reason is that for $h > 0$ the conical singularities of the reduced phase space 
are outside the energy surface.

The non-critical energy surfaces in the reduced phase space are just circles, 
which can best be seen from the Hamiltonian \eqref{SympHam}.
In the critical cases they are simply points.
\end{proof}

The above theorem confirms that inside the image there are no critical points at all.
The momentum map can be considered as a slice through the energy-momentum 
map for fixed energy. The complete image of the energy-momentum map is 
a cone over the polygon. Since the geodesic flow does not change with energy 
(up to scaling) the energy can be fixed to some positive value.
In the following we are interested in the fibres of the energy-momentum map
(for that fixed energy), not just the fibres of the momentum map.

\subsection{The Liouville Foliation}
The fibres over a regular value of the energy momentum map is a $T^3$.  
The fibres of singular values are described by the following theorem.
\begin{theorem} \label{thm:SFibs2}
Points on the edges of the convex polygon are corank one elliptic singular points and have have fibres $T^2$.  The corners correspond to corank two elliptic-elliptic singular points and have fibres $S^1$.
\end{theorem}
\begin{proof}
Consider the corner of the polygon given by $J_1^2 = 2\alpha_1h$, $J_2=0$.  Here the invariants have the values $\pi_1^1 = \alpha_1$, $\pi_2^1 = 2h$, $\pi_3^1 = 0$, $\pi_1^2 = 0$, $\pi_2^2 = 0$, $\pi_3^2 = 0$.  Take a point in the pre-image of the reduction map $R$ from full phase space to reduced phase space
\begin{equation}
R : T^*\mathbb{R}^4 \rightarrow (\pi_1^1,\pi_2^1,\pi_3^1,\pi_1^2,\pi_2^2,\pi_3^2)
\end{equation}
e.g.\ $x = (\sqrt{\alpha_1}, 0, 0, 0)$, $y = (0, \sqrt{2h}, 0, 0)$.  
For the  $SO(2) \times SO(2)$ group action the orbit is
\begin{equation}
\Psi(x, y; \theta_1, \theta_2) = (\tilde x, \tilde y)
\end{equation}
where
\begin{equation}
\tilde x = (\sqrt{\alpha_1} \cos \theta_1, \sqrt{\alpha_1} \sin \theta_1, 0, 0), \quad
\tilde y = (-\sqrt{2h} \sin \theta_1, \sqrt{2h} \cos \theta_1 , 0, 0)
\end{equation}
which is a $S^1$.  Similarly for each of the other corners the fibre is a $S^1$.

The fibres over points on the edge of the polygon are found in a similar manner.  Let $(j_1, j_2)$ be a point in the bifurcation diagram on the edge of the polygon, namely a solution of \eqref{eqn:convexpolygon}, but not a corner.  Then values of $\pi_1^1$, $\pi_2^1$ and $\pi_3^1$ are given by \eqref{eqn:soln1} and \eqref{eqn:soln2}, and the other invariants then follow from the Casimirs and energy equation.  In the reduced space the invariants for a point $(j_1, j_2)$ on the boundary are
\begin{equation}
\pi_1^1 = \alpha_1(1 - \frac {|j_2|} {\sqrt{2h\alpha_2}}), \quad \pi_2^1 = 2h(1 - \frac {|j_2|} {\sqrt{2h\alpha_2}}),
\quad
\pi_1^2 = \sqrt{\frac {\alpha_2} {2h}} |j_2|, \quad \pi_2^2 = \sqrt{\frac {2h} {\alpha_2}} |j_2|
\end{equation}
and $\pi_3^1=0$, $\pi_3^2=0$.
A point in the pre-image of the reduction map is then found, for example,
\begin{equation}
x = ( \sqrt{\pi_1^1}, 0, 0, \sqrt{ \pi_1^2}), \quad
y = ( 0, \sqrt{ \pi_2^1}, \sqrt{ \pi_2^2} ,0) \,.
\end{equation}
Note that as the point $(j_1, j_2)$ is not a corner then none of the values of $x_1$, $x_4$, $y_1$ or $y_3$ can be zero.  Thus the group action of $SO(2) \times SO(2)$ on this point gives $T^2$ as an orbit.

We show the non-degeneracy of the critical points as follows.  Consider first of all the corners of the convex polygon.  We choose the corner for which $J_1=0, J_2=2\alpha_2h$ to show non-degeneracy, but an analogous argument holds for the other corners.  For this corner we have the critical points $x_0=x_1=y_0=y_1=0$.  Here we find $G_1=0, G_2=2h$.  Evaluating the gradients we have $\nabla J_1=0$ and $\nabla G_1=0$, hence the critical points are of corank two.  The Jacobians of the flow for $J_1$ and $G_1$ are then calculated, i.e. $DX_{J_1}$ and $DX_{G_1}$ respectively.  We evaluate the flows for a single point given by $x_2=0, x_3=\sqrt{\alpha_2}, y_3=0, y_2=\sqrt{2h}$, and find $\mu DX_{J_1} + \nu DX_{G_1}$.  This spans the Cartan sub-algebra; the 4 eigenvalues are $\pm \imag \left(\mu \pm 2\nu \sqrt{2h\alpha_1}/ (\alpha_2-\alpha_1) \right)$.  Hence the critical points are of elliptic-elliptic type.

For the edges of the convex polyhedron, excluding the corners, we proceed as follows.  Consider first of all the edge for which $J_1>0, J_2>0$.  We will give the proof of non-degeneracy for this case, but analogous arguments hold for the other edges.  Consider the new integral $K$ which we define by
\begin{equation}
K = \frac {J_1} {\sqrt{\alpha_1}} + \frac {J_2} {\sqrt{\alpha_2}} - \sqrt{2H}.
\nonumber
\end{equation}
This is an integral because $J_1$, $J_2$ and $H$ are integrals.  Notice that $K=0$ is just the equation for the edge of the polyhedron.  First we parameterise the edge, for example choose
\begin{equation}
(j_1,j_2) = (\sqrt{2h\alpha_1}\cos^2(\psi),\sqrt{2h\alpha_2}\sin^2(\psi)),
\end{equation}
for an angle $\psi \in [0,\pi/2]$.

\noindent
Then we find a point in the pre-image of the reduction map, e.g.
\begin{equation}
x_0=\sqrt{\alpha_1}\cos(\psi), \qquad x_1=0, \qquad x_2=\sqrt{\alpha_2}\sin(\psi), \qquad x_3=0,
\nonumber
\end{equation}
\begin{equation}
y_0=0, \qquad y_1=\sqrt{2h}\cos(\psi), \qquad y_2=0, \qquad y_3=\sqrt{2h}\sin(\psi).
\nonumber
\end{equation}
Assume further that $\psi \ne k\pi/2$ for $k \in \Z$ otherwise the point will correspond to a corner of the polyhedron.  Calculating the flow of $K$ we get $X_K=0$. We are on the boundary of the image of the momentum map, the equation of which is $K=0$.  Here we have a corank one critical point.  The Jacobian of the flow $DX_K$ has two non-zero eigenvalues $\pm \imag 2\sqrt{2}/(\alpha_2+\alpha_1+(\alpha_2-\alpha_1)(\frac {j_1} {\sqrt{2h\alpha_1}} - \frac {j_2} {\sqrt{2h\alpha_2}}))$.  These eigenvalues never vanish and are of elliptic type.

We could equally use general theory~\cite{bolsinov} to find the singular fibres on the boundary of the convex polyhedron, as for a corank $r$ singular point of elliptic type the singular fibre is $T^{3-r}$.
\end{proof}

\subsection{Actions}
Since the system is Liouville integrable, we may consider action angle variables in the neighbourhood of any regular point.  Two of the action variables are the angular momenta $J_1$ and $J_2$.  The third action is the action of the reduced system with one degree of freedom
obtained from the Hamiltonian \eqref{SympHam} along the contours $C$ as seen
in Fig.~\ref{fig:Action22}.
Hence we have the following theorem
\begin{theorem} {\em Action Integral.} \label{thm:act2}
The action variables for the geodesic flow on the three dimensional ellipsoid with $SO(2) \times SO(2)$ symmetry are the angular momenta $J_1$, $J_2$ and a third integral $I$ defined by
\begin{equation}
I = \frac {1} {2\pi} \oint_C p_\phi d\phi
\end{equation}
where 
\begin{equation}
p_\phi^2 = \left(2h - \frac {j_1^2} {\alpha_1\cos^2(\phi)} - \frac {j_2^2} {\alpha_2\sin^2(\phi)} \right)\left(\alpha_1\sin^2(\phi)+\alpha_2\cos^2(\phi) \right).
\end{equation}
\end{theorem}
The action $I$ is ``natural" in the technical sense of \cite{dullin02}, and simply means that we consider a cycle $C$ such that $d\theta_1=d\theta_2=0$, where $\theta_i$ is the angle coordinate conjugate to the angular momentum $J_i$. Since the action is natural it is not necessarily smooth.

\begin{figure}
\centerline{\includegraphics[width=8cm]{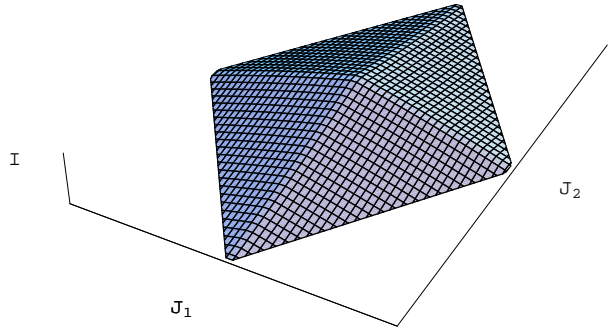}}
\caption{Energy surface in action space with $\alpha_1=2, \alpha_2=2, h=1$.}
\label{fig:engsurf}
\end{figure}
\begin{figure}
\centerline{\includegraphics[width=8cm]{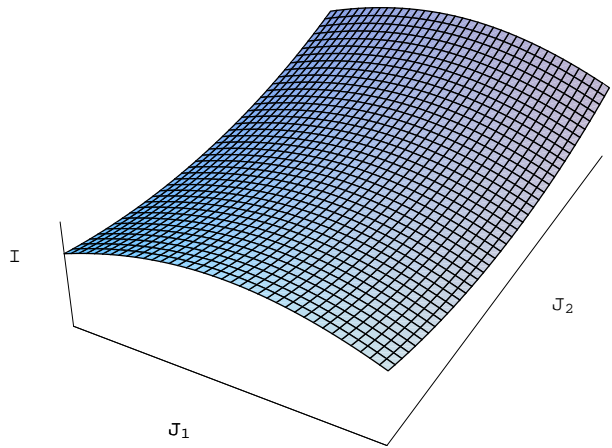}}
\caption{Smooth energy surface in action space with $\alpha_1=2, \alpha_2=2, h=1$.}
\label{fig:smengsurf}
\end{figure}

The energy surface in action space is shown in figure \ref{fig:engsurf}.  Notice that the action $I$ is not differentiable with respect to $J_1$ and $J_2$ on the axes $j_1=0$ and $j_2=0$.  We will prove this by calculating the derivatives of the actions and looking what happens on the axes.  Further note that the faces of the ``pyramid" are not planes. We can smooth the energy surface by plotting instead $I + |J_1| + |J_2|$. This is shown in figure \ref{fig:smengsurf}.

The action integral has simple poles at $\phi=0,\pm \pi$ with residues $\imag j_k$, respectively.  
It has branch points at values of $\phi$ for which the momentum $p_\phi$ vanishes.  
The integral is put into standard form by the substitution $s=\sqrt{\alpha_1}\cos(\phi)$.  
The part of the integrand involving the angular momenta is the square root of the following expression
\begin{equation}
G(s^2) = - \left(s^4 + \frac {\alpha_1j_2^2 - \alpha_2j_1^2 - 2h\alpha_1\alpha_2} {2h\alpha_2} s^2 + \frac {j_1^2\alpha_1} {2h} \right)
\end{equation}
where $G$ factorises as
\begin{equation}
G(s^2) = -(s^2 - s_1^2)(s^2 - s_2^2).
\end{equation}
The polynomial $G$ has $s_1^2$ and $s_2^2$ as roots and hence $\pm s_i$ are branch points 
(in $s$ coordinates) of the action integral.  In $\phi$ coordinates we have branch points at 
$\phi_i$ given by $\alpha_1\cos^2(\phi_i)=s_i^2$.

\begin{figure}
\centerline{\includegraphics[width=8cm]{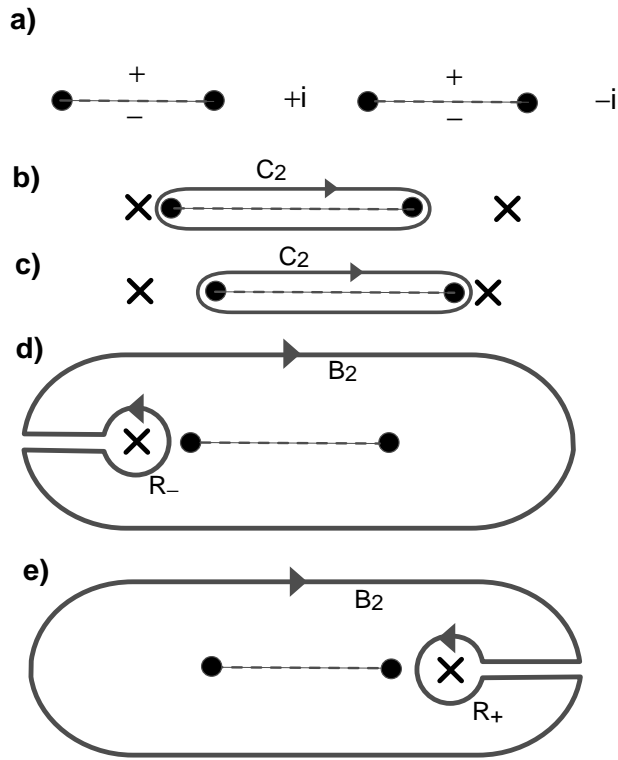}}
\caption{Complex plane $\mathbb{C}(s)$ and choice of branch cuts (a), integration paths for $p_\theta$ for $s_1^2\rightarrow 0$ (b) and $s_2^2\rightarrow \alpha_1$ (c).  (d) and (e) show decomposition of $C_1$ for case (b) and $C_2$ for case (c).}
\label{fig:fig7}
\end{figure}

The action $I$ is a complete elliptic integral with Legendre normal form
\begin{eqnarray}
I & = & \frac {4\sqrt{2h}} {2\pi\sqrt{\alpha_1}} \left(\left(\alpha_1^2+\left(\alpha_2-\alpha_1\right)s_2^2\right)^{1/2}{\cal E}\left(k\right) - \frac {\left(\alpha_1^2+\left(\alpha_2-\alpha_1\right)s_1^2\right)s_2^2} {\alpha_1\left(\alpha_1^2+\left(\alpha_2-\alpha_1\right)s_2^2\right)^{1/2}} \Pi\left(\alpha^2,k\right) \right.
\nonumber \\
& + & \left. \frac {\left(s_2^2 - \alpha_1\right)\left(\alpha_1^2+\left(\alpha_2-\alpha_1\right)s_1^2\right)} {\alpha_1\left(\alpha_1^2+\left(\alpha_2-\alpha_1\right)s_2^2\right)^{1/2}} \Pi\left(\beta^2,k\right)\right)
\end{eqnarray}
where ${\cal E}$ and $\Pi$ are Legendre's complete elliptic integrals of the second and third kind respectively and the parameters are
\begin{equation}
k^2 = \frac {(\alpha_2-\alpha_1)(s_2^2 - s_1^2)} {\alpha_1^2 + (\alpha_2 - \alpha_1)s_2^2}, \quad 
\alpha^2 = - \frac {\alpha_1^2} {s_1^2  (\alpha_2 - \alpha_1)} k^2, \quad
\beta^2 = \frac {\alpha_1\alpha_2} {(\alpha_1 - s_1^2)(\alpha_2 - \alpha_1)} k^2.
\end{equation}
The derivatives of the action $I$ with respect to the other action variables $J_1$ and $J_2$ are
\begin{equation}
\frac {\partial I} {\partial J_1} = - \frac {J_1} {2\pi\alpha_1} \oint_{\gamma} \sqrt{\frac {\alpha_1\sin^2(\phi)+\alpha_2\cos^2(\phi)} {2h - \frac {J_1^2} {\alpha_1\cos^2(\phi)} - \frac {J_2^2} {\alpha_2\sin^2(\phi)}}} \frac {d\phi} {\cos^2(\phi)} = \oint_{\gamma} \beta_1,
\end{equation}
\begin{equation}
\frac {\partial I} {\partial J_2} = - \frac {J_2} {2\pi\alpha_2} \oint_{\gamma} \sqrt{\frac {\alpha_1\sin^2(\phi)+\alpha_2\cos^2(\phi)} {2h - \frac {J_1^2} {\alpha_1\cos^2(\phi)} - \frac {J_2^2} {\alpha_2\sin^2(\phi)}}} \frac {d\phi} {\sin^2(\phi)} = \oint_{\gamma} \beta_2.
\end{equation}
$\frac {\partial I} {\partial J_1}$ has a simple pole at $\phi=\frac {\pi} {2}$ whereas $\frac {\partial I} {\partial J_2}$ has a simple pole at $\phi=0$.  A picture of the complex plane with the branch points and poles is shown in figure \ref{fig:fig7}.
For $J_1=0$ we have $s_1=0$ and $0 \le s_2^2 \le \alpha_1$, and for $J_2=0$ we have $s_2^2 = \alpha_1$ and $0 \le s_1^2 \le \alpha_1$.  In other words, when $J_1=0$ the branch point at $s=s_1$ coincides with the simple pole at $s=0$ (i.e at $\phi=\frac {\pi} {2}$) and when $J_2=0$ the branch point at $s=s_2$ coincides with the simple pole at $s= \alpha_1^{1/2}$ (i.e. at $\phi=0$).  The integration paths in the complex plane are deformed so that the integral may be split up into two separate integrals for each case.  The integral $C_i$ around the branch points is expanded into a loop $B_i$ around the pole in question and then the contribution from the pole $R_i$ is subtracted.  For the case where $s_1$ coincides with $0$ we have
\begin{equation}
\oint_{C_1} = \oint_{B_1} + \oint_{R_1}
\end{equation}
and for the case where $s_2$ coincides with $\alpha_1^{1/2}$ we have
\begin{equation}
\oint_{C_2} = \oint_{B_2} + \oint_{R_2}.
\end{equation}
Evaluating the residues of the integrand at the simple pole in question, we have
\begin{equation}
\res \beta_1 = \frac {J_1} {2\pi i |J_1|}
\text{ and }
\res \beta_2 = - \frac {J_2} {2\pi i |J_2|}.
\end{equation}
We then have
\begin{equation}
\lim_{J_1 \rightarrow 0} \frac {\partial I} {\partial J_1} = \lim_{J_1 \rightarrow 0}  \frac {J_1} {|J_1|}
\text{ and }
\lim_{J_2 \rightarrow 0} \frac {\partial I} {\partial J_2} = \lim_{J_2 \rightarrow 0} - \frac {J_2} {|J_2|}
\end{equation}
by the method of residues.  Hence
\begin{equation}
\lim_{J_1 \rightarrow 0^+} \frac {\partial I} {\partial J_1} = 1, \qquad \lim_{J_1 \rightarrow 0^-} \frac {\partial I} {\partial J_1} = -1, \qquad \lim_{J_2 \rightarrow 0^+} \frac {\partial I} {\partial J_2} = -1, \qquad \lim_{J_2 \rightarrow 0^-} \frac {\partial I} {\partial J_2} = 1 \,.
\end{equation}
Denote the natural actions in each of the four quadrants of the plane in $(J_1,J_2)$ as $I_{++}$, $I_{-+}$, $I_{--}$ and $I_{+-}$.  They are related by
\begin{equation}
\begin{aligned}
I_{-+}(-J_1,J_2) &= S_1I_{++}(J_1,J_2), \\
I_{--}(-J_1,-J_2) &= S_2I_{-+}(-J_1,J_2), \\
I_{+-}(J_1,-J_2) &= S_1I_{--}(-J_1,-J_2), \\
I_{++}(J_1,J_2) &= S_2I_{++}(J_1,-J_2),
\end{aligned}
\end{equation}
where
\begin{equation}
S_1 = \left(
\begin{array}{ccc}
-1 & 0 & 0 \\
0 & 1 & 0 \\
0 & 0 & 1
\end{array}
\right), \qquad
S_2 = \left(
\begin{array}{ccc}
1 & 0 & 0 \\
0 & -1 & 0 \\
0 & 0 & 1
\end{array}
\right) \,.
\end{equation}
We then define unimodular matrices $M_i$ such that the appropriate natural actions and products of actions and $M_i$ join smoothly at the $J_1$  and $J_2$ axes.  We have
\begin{equation}
\begin{aligned}
I_{++} = M_1I_{-+} = M_1S_1I_{++} = M_1I_{++},  & \qquad J_1=0, J_2 > 0, \\
I_{+-} = M_2I_{--} = M_2S_1I_{+-} = M_2I_{+-},  & \qquad J_1=0, J_2 < 0, \\
I_{-+} = M_3I_{--} = M_3S_2I_{-+} = M_3I_{-+},  & \qquad J_2=0, J_1 < 0, \\
I_{++} = M_4I_{+-} = M_4S_2I_{++} = M_4I_{++},  & \qquad J_2=0, J_1 > 0,
\end{aligned}
\end{equation}
on the appropriate axes.

We see that $M_1$ and $M_2$ have an eigenvalue $1$ with eigenvector $(0, J_2, I)^t$ and $M_3$ and $M_4$ have an eigenvalue $1$ with eigenvector $(J_1, 0, I)^t$.  The eigenvector equations and the fact that $M_i \in SL(3,\mathbb{Z})$ show that $M_i$ must have the form
\begin{equation}
M_i = \left(
\begin{array}{ccc}
1 & 0 & 0 \\
\kappa_i & 1 & 0 \\
\beta_i & 0 & 1
\end{array}
\right), i \in \{1,2\}, \qquad
M_i = \left(
\begin{array}{ccc}
1 & \kappa_i & 0 \\
0 & 1 & 0 \\
0 & \beta_i & 1
\end{array}
\right), i \in \{3,4\}.
\end{equation}
In fact, we have $M_1=M_2$ and $M_3=M_4$.  As the derivatives of the actions must join smoothly at the axes we have
\begin{equation}
M_1\frac {\partial I_{-+}} {\partial J_1} = \frac {\partial I_{++}} {\partial J_1}, \qquad M_3\frac {\partial I_{--}} {\partial J_2} = \frac {\partial I_{-+}} {\partial J_2},
\end{equation}
which implies that
\begin{equation}
M_1 = M_2 = \left(
\begin{array}{ccc}
1 & 0 & 0 \\
0 & 1 & 0 \\
2 & 0 & 1
\end{array}
\right), \qquad
M_3 = M_4 = \left(
\begin{array}{ccc}
1 & 0 & 0 \\
0 & 1 & 0 \\
0 & -2 & 1
\end{array}
\right).
\end{equation}
Taking a loop around around the origin and calculating the overall monodromy matrix $M$, gives 
\begin{equation}
M = (M_4S_2)^{-1}(M_2S_1)^{-1}(M_3S_2)(M_1S_1) = \id,
\end{equation}
so monodromy is not present in the system.  Globally smooth action variables can be defined by $I$, $M_1S_1I$, $M_3S_2M_1S_1I$ and $(M_2S_1)^{-1}M_3S_2M_1S_1I$ in each of the four quadrants of the plane, where $I$ is the natural action in the $(+,+)$ plane. 
Thus we have proved
\begin{theorem}
The geodesic flow on the three dimensional ellipsoid with two pairs of equal axes 
has three global actions.
\end{theorem}

Two of these actions are trivial, they are the generators of the symmetry. 
The third action is given by a non-trivial complete elliptic integral,
and when using the separating coordinate system this action is not globally smooth.
Instead it is only continuous across the lines $J_k = 0$. Nevertheless, even though this
natural action is not differentiable, different linear combinations exist which give 
actions that are smooth across the lines $J_k = 0$. Moreover, completing a cycle 
around the origin there is no overall monodromy. Therefore a globally smooth and 
single valued third action exists in this case. It would be interesting to check whether 
it is possible to find a single formula in terms of standard complete elliptic integrals 
which is smooth.

The absence of monodromy was to be expected since there are no critical 
points inside the image of the momentum maps. Nevertheless, from the point 
of view of the explicit calculation of the third momentum this is a non-trivial 
observation.

\section{Ellipsoid with 112 symmetry}
Consider the geodesic flow on a three ellipsoid with the two largest semi-axes equal, namely $0 < \alpha_0 < \alpha_1 < \alpha_2 = \alpha_3$, corresponding to the symmetry group action $SO(1) \times SO(1) \times SO(2)$.  The action of $SO(1)$ has no effect as it merely fixes the variables on which it acts, but we include it in a direct product to label the case where we have two equal semi-axes so as to distinguish whether it is the two largest, the middle two or the two smallest semi-axes which are equal. We will call the 112 ellipsoid the three dimensional oblate ellipsoid, in analogy to the two dimensional ellipsoid of revolution with a $SO(1) \times SO(2)$ symmetry, which is oblate or disc shaped.  The Casimirs $C_1$ and $C_2$, the resulting Dirac bracket \eqref{eqn:dirac} and the Hamiltonian are the same as in the generic case, with $\alpha_2$ and $\alpha_3$ set equal.  In this situation the integrals $F_2$ and $F_3$ from the generic case are not defined any more, but the singular terms cancel in the sum $G = F_2 + F_3$.  The other integrals $F_0$ and $F_1$ remain the same. 
The system is invariant under rotations in the $(x_2, x_3)$ plane and its cotangent lift, the $SO(2)$ group action being
\begin{equation}
\Phi(x, y; \theta) = (\tilde x, \tilde y)
\end{equation}
where
\begin{equation}
\begin{aligned}
\tilde x &= (x_0, x_1, x_2\cos \theta-x_3\sin \theta, x_2\sin \theta+x_3\cos \theta) \\
\tilde y &= ( y_0, y_1, y_2\cos \theta-y_3\sin \theta, y_2\sin \theta+y_3\cos \theta) \, .
\end{aligned}
\end{equation}
The group action $\Phi$ is the flow generated by the angular momentum $J = x_2 y_3 - x_3 y_2$, which is a global action variable.

\subsection{Liouville Integrability}
\begin{theorem} {\em Liouville Integrability.} \label{thm:ALInt112}
The Geodesic flow on the ellipsoid with equal largest semi-axes is Liouville integrable. Constants of motion are the energy, $H=\frac {1} {2} \left(y_0^2+y_1^2+y_2^2+y_3^2\right)$, 
the angular momentum, $J = x_2y_3 - x_3y_2$, 
and the third integral $G = F_2 + F_3$
\begin{equation}
G = y_2^2 + y_3^2 + \frac {\left(x_0y_2 - x_2y_0\right)^2} {\alpha_2 - \alpha_0} + \frac {\left(x_0y_3 - x_3y_0\right)^2} {\alpha_2 - \alpha_0} + \frac {\left(x_1y_2 - x_2y_1\right)^2} {\alpha_2 - \alpha_1} + \frac {\left(x_1y_3 - x_3y_1\right)^2} {\alpha_2 - \alpha_1}
\end{equation}
\end{theorem}

\begin{proof}
The relationship between the integrals and the Hamiltonian is $2 H = F_0 + F_1 + G$, and in the limit $\alpha_3 \to \alpha_2$ the constant of motion $(\alpha_2 - \alpha_3)F_2$ becomes $J^2$.  Hence $G$ and $H$ commute because the $F_i$ commute in the generic case, and $J^2$ commutes with $H$ and $G$, and hence so does $J$.  Here the relationship between the constants of motion on the symplectic leaves of the Dirac bracket is given by
\begin{equation} \label{FGrel112}
   \frac{F_0}{\alpha_0} + \frac{F_1}{\alpha_1} + \frac{G}{\alpha_2} - \frac{J^2}{\alpha_2^2} = 0 \,,
\end{equation}
which is the limit of the generic relation $\sum F_i/\alpha_i = 0$
using $F_2/\alpha_2 + F_3/\alpha_3 = G/\alpha_3 - F_2(\alpha_2 - \alpha_3)/(\alpha_2 \alpha_3)$.

$H$, $J$, $G$ and the Casimirs $C_1$, $C_2$ are functionally independent almost everywhere on the level set $C_1 = C_2 =0$: They are polynomial and independent e.g. at $x=(\sqrt{\alpha_0}, 0, 0, 0)$, $y=(0, 0, \sqrt{2h}, 0)$.
\end{proof}

The group action $\Phi$ has the invariants
\begin{equation}
\pi_1 = x_2^2+x_3^2,\quad
\pi_2 = y_2^2+y_3^2,\quad
\pi_3 = x_2y_2 +x_3 y_3,\quad
\pi_4 = x_2 y_3 -x_3 y_2\,,
\end{equation}
related by $\pi_1\pi_2 -\pi_3^2 -\pi_4^2 =0$. The remaining variables $x_0,x_1,y_0,y_1$
are trivial invariants of $\Phi$. The only fixed point of $\Phi$ is $x_2=x_3=y_2=y_3=0$.
As before, when $J=\pi_4 = j \not = 0$ this fixed point is not in $J^{-1}(j)$ and the reduction by
the $SO(2)$ symmetry leads to a smooth reduced system on $J^{-1}(j)/SO(2)$:

\begin{lemma}  \label{thm:Regredn112}
A set of reduced coordinates $(\xi_0, \xi_1, \xi_2, \eta_0, \eta_1, \eta_2)$ is defined on the reduced phase space $P_j = J^{-1}(j)/SO(2)$ by the formulae
\[
   \xi_0=x_0, \quad \xi_1=x_1, \qquad \xi_2=\sqrt{\pi_1}, \qquad
   \eta_0=y_0,\quad \eta_1=y_1, \qquad \eta_2 =  \frac{\pi_3}{\sqrt{\pi_1}}.
\]
The reduced coordinates satisfy the Dirac bracket in $ \R^6[\xi,\eta]$, i.e. $\{.,.\}_6$ as defined in \eqref{eqn:dirac}.
The mapping $R:\R^8[x,y] \to \R^6[\xi,\eta]$ is Poisson from $\R^8$ with $\{.,.\}_8$ to $\R^6$ with $\{.,.\}_6$ and the reduced system has reduced Hamiltonian
\[
\hat H = \frac12 ( \eta_0^2 +\eta_1^2+\eta_2^2)+ \frac{j^2}{2\xi_2^2} 
\]
and additional integral
\[
\hat G = \eta_2^2 + \frac{(\xi_2\eta_0 - \xi_0 \eta_2)^2}{\alpha_2 - \alpha_0} 
                            + \frac{(\xi_2\eta_1 - \xi_1 \eta_2)^2}{\alpha_2 - \alpha_1}
      + \frac{j^2}{\xi_2^2} \left( 1 +  \frac{\xi_0^2}{\alpha_2 - \alpha_0} 
                                                        + \frac{\xi_1^2}{\alpha_2 - \alpha_1} \right)  \,.
\]
\end{lemma}

\begin{proof}
Define a set of coordinates on $\R^6[\xi,\eta]$ as shown.
The Poisson property of the map $R$, i.e. $\{ f \circ R, g \circ R\}_8 = \{f, g\}_6 \circ R$
follows from direct computation of the basic brackets, e.g.
$\{ \xi_1, \xi_2 \}_6 = \{ x_1, \sqrt{x_2^2+x_3^2} \}_8 =0$, etc. The reduced Hamiltonian $\hat H$ and additional integral $\hat G$ are found by writing them in terms of the invariants and then expressing them in terms of the reduced variables.
\end{proof}

The reduced system is the ``geodesic flow" on the 2-dimensional ellipsoid
with semi-axes $\sqrt{\alpha_0}, \sqrt{\alpha_1}, \sqrt{\alpha_2}$ and an additional
effective potential $j^2/2\xi_2^2$, and as $\xi_2 > 0$, the reduced system for $|j|>0$ has only the open half of the ellipsoid as configuration space with the plane $\xi_2=0$ dynamically not accessible.

We define a singular coordinate system on $\R^8[x_i,y_i]$ by
\begin{equation}
x_2 = \xi_2 \cos\theta, \qquad x_3 = \xi_2 \sin\theta
\end{equation}
where $\theta$ is the angle of rotation corresponding to the $SO(2)$ symmetry group action $\Phi$. 
We then choose confocal ellipsoidal coordinates and parameterise the 2-ellipsoid to give us a set of generalised coordinates $(\lambda_1, \lambda_2, \theta)$ on the three ellipsoid with the larger two semi-axes equal.  We denote the conjugate momenta are denoted by $(p_1, p_2, p_\theta)$.

\begin{lemma} \label{thm:Separation112}
The Hamiltonian for the geodesic flow on the ellipsoid with equal largest semi-axes in local symplectic coordinates reads
\begin{eqnarray}
H & = & - \frac {2(\alpha_0 - \lambda_1)(\alpha_1 - \lambda_1)(\alpha_2 - \lambda_1)} {\lambda_1(\lambda_2 - \lambda_1)}p_1^2 
- \frac {2(\alpha_0 - \lambda_2)(\alpha_1 - \lambda_2)(\alpha_2 - \lambda_2)} {\lambda_2(\lambda_1 - \lambda_2)} p_2^2 \nonumber \\ 
& + & \frac {(\alpha_0-\alpha_2)(\alpha_1-\alpha_2)} {2\alpha_2(\lambda_1 - \alpha_2)(\lambda_2 - \alpha_2)} p_\theta^2. 
\nonumber
\end{eqnarray}
The constants of motion are $p_\theta$ and $\tilde G_i$ 
\begin{equation}
\tilde G_i = \frac {2(\alpha_0 - \lambda_i)(\alpha_1 - \lambda_i)(\alpha_2 - \lambda_i)} {\lambda_i} p_i^2 - h\lambda_i - \frac {(\alpha_0-\alpha_2)(\alpha_1-\alpha_2)} {2\alpha_2(\lambda_i - \alpha_2)} p_\theta^2 \nonumber
\end{equation}
where $i =1,2$.  The integrals $G$ and $\tilde G_i$ 
are related by
\begin{equation}
\tilde G_1 + \tilde G_2 = \frac {(\alpha_2-\alpha_0)(\alpha_2-\alpha_1)G} {\alpha_2} -2\alpha_2h + \frac {\alpha_2^2 -\alpha_0\alpha_1} {\alpha_2^2}p_\theta^2. \nonumber
\end{equation}
\end{lemma}

\begin{proof}
The Hamiltonian is found by applying the cotangent lift to the new coordinates.   The variables are separated by multiplication by $\lambda_2-\lambda_1$ and rearranging to find the $\tilde G_i$.

By separation of variables the momenta $p_k$ conjugate to $\lambda_k$ can 
be expressed as
\begin{equation}
   p_k^2 = -\frac{\tilde Q( \lambda_k)}{4 A(\lambda_k)}
\end{equation}
where $\tilde Q$ is a cubic polynomial.
\end{proof}

The relation of the separation constant $\tilde G$ to the constant of motion $G$
is
\[
\frac{F_0}{z-\alpha_0}+ \frac{F_1}{z-\alpha_1}+ \frac{G}{z-\alpha_2}+ \frac{J^2}{(z-\alpha_2)^2}
= 
\frac{\tilde Q(z)} {A(z)} \,.
\]

\subsection{Singular Reduction}

\begin{lemma} \label{thm:SingRedn112}
The singular reduced phase space of the geodesic flow on the 3-ellipsoid with equal largest axes is the phase space of the geodesic flow on the 2-ellipsoid reduced by the $\Z_2$ action 
$S(\xi_0,\xi_1,\xi_2,\eta_0,\eta_1,\eta_2) = (\xi_0,\xi_1,-\xi_2, \eta_0,\eta_1,-\eta_2)$.
Thus it is the geodesic flow on the 2-ellipsoid with a hard billiard wall inserted
in the $\xi_2=0$-plane.
\end{lemma}

\begin{proof}
Singular reduction for $j=0$ leads to a reduced system on a non-smooth manifold. We use invariant theory to carry out the singular reduction. The reduced phase space is a subset of $\R^7[x_0, y_0, x_1, y_1, \pi_1, \pi_2, \pi_3]$ and is defined by the two Casimirs
\begin{equation}
\frac {x_0^2} {\alpha_0} + + \frac {x_1^2} {\alpha_1} + \frac {\pi_1} {\alpha_2}  = 1, \qquad \frac {x_0y_0} {\alpha_0} + \frac {x_1y_1} {\alpha_1} + \frac {\pi_3} {\alpha_2} = 0,
\label{eqn:Casimirs112}
\end{equation}
the relation between the invariants $\pi_1\pi_2 - \pi_3^2=j^2$  and the inequalities $\pi_1\ge0$, $\pi_2\ge0$.  Since the Casimirs are linear in the invariants we can eliminate them from the relation to give
\begin{equation} \label{eqn:casi112}
\alpha_1\left(1 - \frac {x_0^2} {\alpha_0} - \frac {x_1^2} {\alpha_1}\right)\pi_2 -\alpha_1^2\left(\frac {x_0y_0} {\alpha_0} + \frac {x_1y_1} {\alpha_1}\right)^2 = j^2 \, ,
\end{equation}
which defines a four dimensional object.  This is the reduced phase space $P_j =J^{-1}(j)/SO(2)$ which is a smooth 4 dimensional manifold for $j \ne 0$ and not a smooth object for $j=0$.

The geodesic flow on the 2-ellipsoid reduced by the $\Z_2$ action $S$ is the billiard.  In \cite{BDD} we proved that the singular reduced phase space of the geodesic flow on the 3-ellipsoid with equal middle semi-axes and vanishing angular momentum $j=0$ is the phase space of the geodesic flow on the 2-ellipsoid reduced by a $\Z_2$ action.  For the 112 situation the result is analogous, although here the $\Z_2$ action fixes the plane orthogonal to the largest axis, i.e. $\xi_2=0$, whereas in the 121 case the $\Z_2$ action fixed the plane orthogonal to the middle axis, namely $\xi_1=0$.  The reader is referred to \cite{BDD} for a full description of the proof.
\end{proof}

\subsection{The Liouville foliation}

We investigate the topology of the invariant level sets obtained by fixing the constants of motion.  The energy momentum map is ${\cal EM} = (H, J, G) : M \rightarrow \mathbb{R}^3$.  Since $H$ for a geodesic flow is homogeneous in the momenta we can fix the energy to, say, $h$.

\begin{theorem} \label{thm:EM112}
The image of the energy momentum map ${\cal EM}$ for constant energy $H=h$ is the region in $\mathbb{R}^2$ bounded by the quadratic curves (see figure~\ref{fig:BiDi112})
\begin{equation}
g = \frac {2\alpha_2} {\alpha_2-\alpha_1} h - \frac {\alpha_1} {\alpha_2(\alpha_2-\alpha_1)} j^2,
\label{eqn:bdrycurves112A}
\end{equation}
\begin{equation}
g = \frac {2\alpha_2} {\alpha_2-\alpha_0} h - \frac {\alpha_0} {\alpha_2(\alpha_2-\alpha_0)} j^2,
\label{eqn:bdrycurves112B}
\end{equation}
\begin{equation}
g = \sqrt{\frac {8\alpha_2h} {(\alpha_2-\alpha_0)(\alpha_2-\alpha_1)}} |j| - \frac {\alpha_2^2-\alpha_0\alpha_1} {\alpha_2(\alpha_2-\alpha_0)(\alpha_2-\alpha_1)} j^2.
\label{eqn:bdrycurves112C}
\end{equation}
Singular values of the energy momentum map are the boundary curves (elliptic), 
their transverse intersections (elliptic-elliptic), the boundary between the two chambers of the bifurcation diagram (hyperbolic), and the non-transverse intersections between the boundary curves (degenerate).
\end{theorem}

\begin{proof}
As in the generic case critical points can occur on sub-ellipsoids.
On $x_0 = y_0 = 0$ the integral $F_0 = 0$ and $\nabla F_0 = 0$,
similarly for $x_1 = y_1 = 0$ and $F_1$. In both cases the corresponding
sub-ellipsoids are ellipsoids of revolution. 
The image of the critical points
with $x_0 = y_0 = 0$ is found using the relation \eqref{FGrel112}
to eliminate $F_1$ in $2H = F_0 + F_1 + G$, which gives
\[
    2H = G - \alpha_1\left( \frac{G}{\alpha_2}  - \frac{J^2}{\alpha_2^2} \right),
\]
and hence the upper boundary curve of critical values \eqref{eqn:bdrycurves112A}. 
A similar computation for critical points with $x_1 = y_1 = 0$
gives the lower curve \eqref{eqn:bdrycurves112B}, which divides the bifurcation diagram into two chambers.

Since for $x_i = y_i = 0$ the integrals $F_i = 0$ and
also its gradient vanishes, $G = F_2 + F_3$ and
its gradient clearly vanishes when 
$x_2 = x_3 = y_2 = y_3 = 0$.
Considering the Casimirs the solutions set of $x_2 = x_3 = y_2 = y_3 = 0$ 
is a geodesic flow on the ellipse in the $x_0$-$x_1$ plane. Fixing the
energy two critical circles are obtained.
On these critical points also $J =0$ so that the origin in the image $(j, g) = (0,0)$ 
is a critical value.  This is analogous to the case with equal middle semi-axes.  However in this case the critical value is not isolated but lies on the intersection of the two curves defined by \eqref{eqn:bdrycurves112C}.  The curves \eqref{eqn:bdrycurves112C} arise in the following way.

We use the ellipsoidal coordinates from lemma~\ref{thm:Separation112} to establish which other points are non-singular. 
These coordinates are non-singular outside any sub-ellipsoid $\xi_i = 0$. 
To find critical points in the region of phase space 
with $\xi$ coordinates such that all $\xi_i \not = 0$
we compute the rank of the matrix $D(\tilde G_1, \tilde G_2, p_\theta)$,
see lemma.~\ref{thm:Separation112}. 
Since the variables are separated this implies $p_i = 0$ 
and 
$2h \alpha_2(\lambda_i - \alpha_2)^2 = (\alpha_2-\alpha_0)(\alpha_2-\alpha_1) p_\theta^2 > 0$
so here we have critical points not contained in the coordinate singularities $\xi_i=0$.  Solving for $\lambda_i$ and substituting this value and $p_i=0$ into the expression for $\tilde G_i$ in lemma \ref{thm:Separation112} gives the following curves in $(j,\tilde g)$ coordinates
\begin{equation}
\tilde g = -h\alpha_2 \pm |j|\sqrt{\frac {2h(\alpha_2-\alpha_0)(\alpha_2-\alpha_1)} {\alpha_2}}
\end{equation}
and using the relationship between $g$ and $\tilde g$ in lemma~\ref{thm:Separation112} to convert to $(j,g)$ coordinates we get the curves \eqref{eqn:bdrycurves112C}.  For the ellipsoid with equal middle axes the equation arising here had no solution.  The curve intersects tangentially with the curve corresponding to critical points $x_1 = y_1 = 0$ at the points
\begin{equation}
(j,g) = \left(\pm \sqrt{\frac{2h\alpha_2(\alpha_2-\alpha_1)} {\alpha_2-\alpha_0}},\frac{2h} {(\alpha_2-\alpha_0)^2} (\alpha_2^2-2\alpha_0\alpha_2+\alpha_0\alpha_1)\right)
\label{eqn:tgt}
\end{equation}

Now let us investigate the types of critical points.  For the corank one critical points on the upper boundary curve \eqref{eqn:bdrycurves112A} these points are non-degenerate because the Jacobian $DX_{F_0}$ of the flow generated by $F_0$ restricted to the critical points $x_0 = y_0 = 0$ is given by the following non-zero sub-block with $i=0$ (all other sub-blocks having zero entries at the critical points)
\begin{equation}
\label{DXF0}
\begin{pmatrix}
  -2 K_i(x,y) &  2(K_i(x, x) - 1) \\
  -2K_i(y, y) &  2K_i(x, y)  
\end{pmatrix} \qquad \text{where} \quad
K_i(x, y) = \sum_{k \not = i} \frac{x_k y_k}{\alpha_k - \alpha_i} \,.
\end{equation}
Note that $K_0(y,y)$ never vanishes because $\alpha_k-\alpha_0$ in the denominator of each term is always positive.  The matrix is traceless so the square of the eigenvalues is given by the negative determinant.  Evaluating the matrix on the point $x_0 = x_1 =  x_3 = 0$, $y_0 = y_2 = 0$ shows that the eigenvalues of this matrix never vanish and are of elliptic type, and hence these critical points are non-degenerate.

For the corank one critical points on the lower boundary curve \eqref{eqn:bdrycurves112B} we have the Jacobian of the flow generated by $F_1$ restricted to the critical points $x_1 = y_1 = 0$ as given by \eqref{DXF0}.  However here $K_1(y,y)$ can vanish because the terms in the sum can have different signs.  If we consider the point $x_0 = x_3 = 0$ on all critical sets we have $x_2 = \pm \sqrt{\alpha_2}$ and $y_2=0$.  The eigenvalues of the matrix block vanish at $y_0^2 = 2h(\alpha_1-\alpha_0)/(\alpha_2-\alpha_0)$ so here we have degenerate points.  These corresponds to the two points of tangency where the lower boundary curve \eqref{eqn:bdrycurves112B} and third curves \eqref{eqn:bdrycurves112C} intersect, defined by \eqref{eqn:tgt}.  For values of $j^2 < \sqrt{\frac{2h\alpha_2(\alpha_2-\alpha_1)} {\alpha_2-\alpha_0}}$ we have $y_0^2 > 2h(\alpha_1-\alpha_0)/(\alpha_2-\alpha_0)$ and here the eigenvalues are real and hence hyperbolic.  Note that this is the case even for the point on the axis $j=0$.  For  $j^2 > \sqrt{\frac{2h\alpha_2(\alpha_2-\alpha_1)} {\alpha_2-\alpha_0}}$ we have $y_0^2 < 2h(\alpha_1-\alpha_0)/(\alpha_2-\alpha_0)$ and the eigenvalues are purely imaginary and hence elliptic.

The two corank two points given by the intersection of the two curves \eqref{eqn:bdrycurves112A} and \eqref{eqn:bdrycurves112B} are also non-degenerate, because the non-zero $2\times 2$ blocks
of the Jacobians are distinct, so that $\mu DX_{F_0} + \nu DX_{F_1}$
spans the Cartan subalgebra; the 4 eigenvalues (for any point on the critical circles
given by $x_0 = x_1 = y_0 = y_1 = 0$) are
$\pm2  \imag \mu  \sqrt{2 \alpha_0 h}/(\alpha_2 - \alpha_0)$ and 
$\pm2 \imag \nu \sqrt{2\alpha_1h }/(\alpha_2 - \alpha_1)$.
This orbit is a relative equilibrium, i.e. a circle in the $x_1$-$x_2$ plane.
The eigenvalues of $DX_{F_0}$ and $DX_{F_1}$ are elliptic,
so at their intersection an orbit of elliptic-elliptic type is found.

Now consider the type of singularities on the third curves \eqref{eqn:bdrycurves112C}.  These are corank one critical points.  We consider the Jacobian matrix of the flow generated by the one degree of freedom system corresponding the the constant of motion $\tilde G$.  The eigenvalues $\lambda$ are given by $\lambda^2 = - \frac {\partial ^2 \tilde G} {\partial \lambda_s^2} \frac {\partial ^2 \tilde G}{\partial p_s^2}$ and both partial derivatives are strictly positive.  Hence critical points on these lines are elliptic.  For the origin on these lines we have $J=0$ and $G=0$.  This is a critical point of corank two.  Moreover $\nabla J = 0$ as well,
and the Jacobian of $X_J$ has eigenvalues $\pm \imag$, since  its flow $\Phi$ is a rotation. 
We have that $\mu DX_{G} + \nu DX_{J}$ spans the Cartan subalgebra; the 2 eigenvalues (for any point on the critical circles) are
$ \pm \imag \mu \sqrt{ 8 \alpha_2 h / (\alpha_2-\alpha_0)(\alpha_2-\alpha_1)} \pm \imag \nu$.
Hence the origin is of elliptic-elliptic type.

This establishes the existence, non-degeneracy or degeneracy, and type of all critical points.
The bifurcation diagram is shown in figure~\ref{fig:BiDi112}.

It remains to show that there are no other critical points.  We check the pre-images in full phase space of the sub-ellipsoids $\xi_i = 0$ where the ellipsoidal coordinates are not defined.
Here the argument is that given for the case of the ellipsoid with equal middle semi-axes in \cite{BDD}.

\begin{figure}
\centerline{\includegraphics[width=8cm]{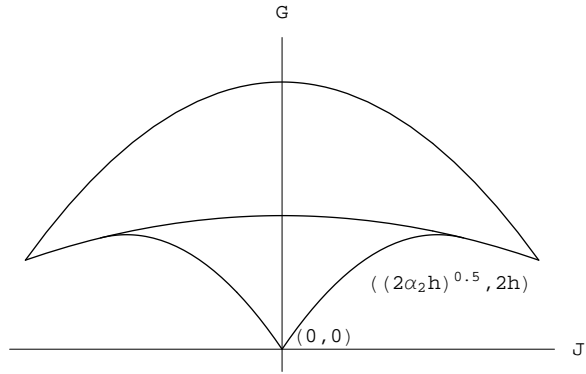}}
\caption{Bifurcation Diagram for the ellipsoid with $\alpha_0 < \alpha_1 < \alpha_2 = \alpha_3$ for $\alpha_0=1, \alpha_1=2, \alpha_2=3, h=1$.}
\label{fig:BiDi112}
\end{figure}

\end{proof}

The bifurcation diagram figure~\ref{fig:BiDi112} can be considered as the square root of the
diagram in figure~\ref{DegenBifDiags} bottom left, similar to what happened in the case with two equal middle semi-axes~\cite{BDD}.

We now find the fibres of the energy momentum map at the singular values in the bifurcation diagram.

\begin{theorem} \label{thm:SingFib112}
The singular fibres over the upper boundary curve \eqref{eqn:bdrycurves112A} of the image of the energy momentum map at constant energy, with the exception of its intersections, are two-dimensional tori $T^2$.  At each intersection point of the upper and lower boundary curves \eqref{eqn:bdrycurves112A} and \eqref{eqn:bdrycurves112B} the singular fibre is $S^1$.  The singular fibre at the origin is two circles $S^1$.  For the boundary curves \eqref{eqn:bdrycurves112C} except at its intersections the singular fibre is two sets of two-dimensional tori $T^2$.  For the lower boundary curve \eqref{eqn:bdrycurves112B} the singular fibre is a two-dimensional torus $T^2$ on the boundary of the bifurcation diagram, $B \times T^2$ on the region which divides the bifurcation diagram into two chambers, and $T^2$ for the two degenerate points where the curve is tangential to \eqref{eqn:bdrycurves112C}, where $B$ is the Bolsinov-Fomenko atom. The multiplicity of the regular fibres $T^3$ is one for the chamber enclosed between the upper boundary curve \eqref{eqn:bdrycurves112A} and the lower boundary curve \eqref{eqn:bdrycurves112B} and two for the chamber enclosed between the lower boundary curve \eqref{eqn:bdrycurves112B} and the curves \eqref{eqn:bdrycurves112C}. 
\end{theorem}

\begin{proof}
Consider first of all the upper boundary curve \eqref{eqn:bdrycurves112A} with $F_0=0$ of the image of the energy momentum map.  On this curve all singularities are of elliptic type, and hence 
the singular fibre is $T^{3-r}$ where $r$ is the corank of the singularity; hence we have $T^2$ on the curve and $S^1$ at its intersection points with the curve \eqref{eqn:bdrycurves112B}.

The curve consists of all orbits in the geodesic flow on the ellipsoid
of revolution defined by $x_0=y_0=0$, which is an oblate two dimensional ellipsoid, obtained by deleting the smallest axis in the three dimensional ellipsoid.  Reduction maps each $T^2$ of this system to a relative periodic orbit.  The isolated periodic orbit in the $23$-plane of the geodesic flow 
on the ellipsoid of revolution corresponds to the extremal points with $J=\pm\sqrt{2\alpha_2 h}$.
Reduction maps this relative equilibrium to the fixed point $\xi = (0, 0, \sqrt{\alpha_2})$ 
on the larger-axis of the reduced ellipsoid.  We also note that the multiplicity of the fibres is one on this boundary.  If we consider the fibres at the intersection points we find that at one intersection we have the periodic orbit in the $23$ plane, in full phase space, in one direction $J=+\sqrt{2\alpha_2 h}$, and at the other the other periodic orbit in the opposite direction $J=-\sqrt{2\alpha_2 h}$, so the multiplicity is one at the intersections.  Moving from the intersection to the boundary curve $F_0=0$ the multiplicity of $T^2$ must therefore also be one.

Consider now the other transverse intersection of boundary curves at $G=0$, $J=0$.  This is an elliptic-elliptic point of corank two and the fibre is $2S^1$. The fibres are the periodic orbits in the $01$ plane on the ellipsoid in full phase space corresponding to $\pi_1=\pi_2=\pi_3=0$.  There are two periodic orbits; one in either direction.

Now consider the curves defined in \eqref{eqn:bdrycurves112C} on which the origin $(J,G)=(0,0)$ lies.  For points on the curves away from the tangential intersection with \eqref{eqn:bdrycurves112B} the critical points are of elliptic type and of corank one.  Hence the singular fibres are $T^2$.  Since the multiplicity at the origin is two, the multiplicity on the boundary curves must also be two.

Finally consider the lower boundary curve \eqref{eqn:bdrycurves112B} corresponding to $F_1=0$.  The curve consists of all orbits in the geodesic flow on the ellipsoid of revolution defined by $x_1=y_1=0$.  This is an oblate two dimensional ellipsoid, as occurred for the curve corresponding to $F_0=0$, but in the current instance the ellipsoid of revolution is obtained by deleting the second smallest axis in the three dimensional ellipsoid.  This is why hyperbolic behaviour is possible on this line, and also why the line can have degenerate points.  For the part of this curve which is a boundary of the bifurcation diagram, excluding the two degenerate points, the critical points are elliptic and corank one, and by the same argument as for the curve \eqref{eqn:bdrycurves112A} the singular fibres are $T^2$ of multiplicity one.

Before looking at the remaining part of \eqref{eqn:bdrycurves112B} let us consider the multiplicity of the regular fibres.  The regular fibres are $T^3$ by the Liouville-Arnold theorem. For the chamber enclosed between the upper boundary curve \eqref{eqn:bdrycurves112A} and the lower boundary curve \eqref{eqn:bdrycurves112B} the multiplicity is one because the multiplicity is one on the boundaries.  For the chamber enclosed between the lower boundary curve \eqref{eqn:bdrycurves112B} and the curves \eqref{eqn:bdrycurves112C} the multiplicity is two as the multiplicity is two on the boundaries.

As we cross the part of the lower boundary curve \eqref{eqn:bdrycurves112B} which divides the bifurcation diagram into two chambers, the multiplicity of the regular fibres changes from one to two.  From the Fomenko theory of atoms we therefore require an atom corresponding to a bifurcation from a multiplicity of one to two.  A likely candidate for the singular fibre in the reduced phase space is the direct product of $S^1$ and the Fomenko atom $B$.  We shall show that this is indeed what we get.  The critical points are of corank one but are of hyperbolic type.  In the reduced ellipsoid the critical sets are circles in the $02$ plane.  However, the singular fibres also have a separatrix because the critical points are hyperbolic.

\begin{figure}
\centerline{\includegraphics[width=6cm]{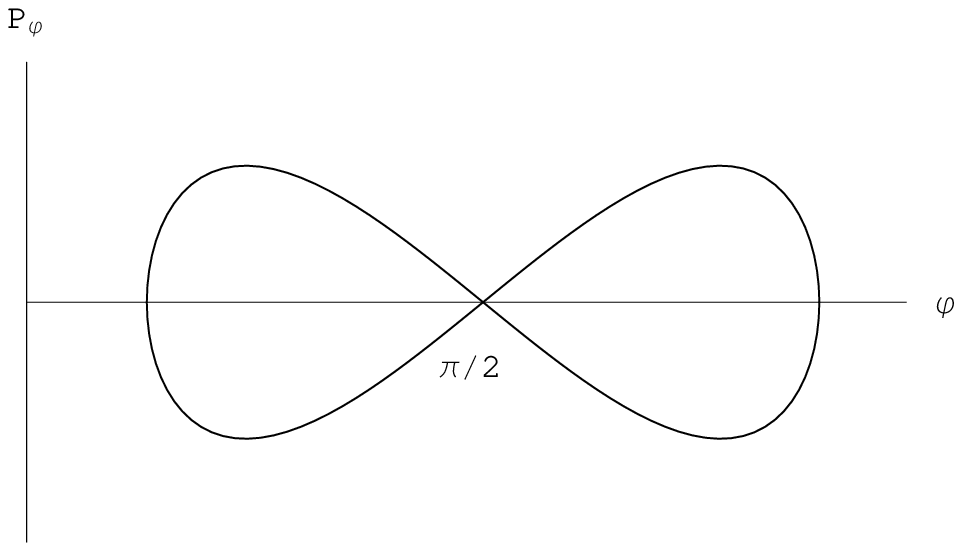}}
\caption{The intersection of the preimage of points on \eqref{eqn:bdrycurves112B} for $j \ne 0$ with the Poincare section $x_0=0$ in reduced phase space. The separatrix is of type B.}
\label{fig:fig2112a}
\end{figure}

\begin{figure}
\centerline{\includegraphics[width=6cm]{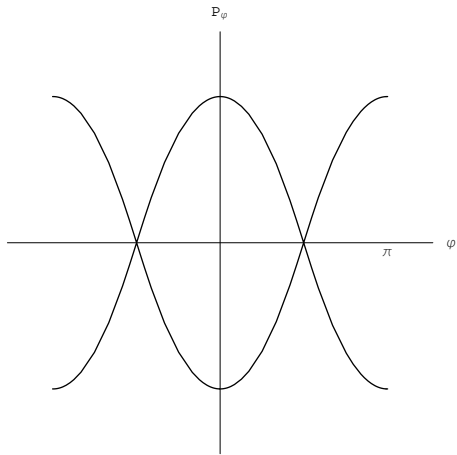}}
\caption{The intersection of the preimage of points on \eqref{eqn:bdrycurves112B} for $j = 0$ with the Poincare section $x_0=0$ in reduced phase space. The separatrix is of type $C_2$.}
\label{fig:fig2112b}
\end{figure}

By lemma~\ref{thm:SingRedn112} the reduced system for $j=0$ is the geodesic flow on the 2-ellipsoid 
quotient by the $\Z_2$ action $S$.  Ignoring the quotient the reduced singular fibre consists 
of the unstable periodic orbits corresponding to $\xi_1=\eta_1=0$ on the reduced ellipsoid and their separatrix.
This can be seen from the Poincar\'e section $\xi_0=0$. Since $\xi_0 = \eta_0 =0$ 
is an invariant subflow the boundary of the section with $\eta_0 \ge 0$ is an 
invariant set and it is the only place where the flow is not transverse to the section.
In configuration space the section condition is the ellipse in the $12$-plane,
and it can be parametrised by an angle $\phi$ by
$(\xi_1,\xi_2) = ( \sqrt{\alpha_1} \cos\phi, \sqrt{\alpha_2} \sin\phi )$.
The momentum $p_\phi$ conjugate to $\phi$ then gives the momenta as
$(\eta_1,\eta_2) = ( -\sqrt{\alpha_1} \sin\phi, \sqrt{\alpha_2} \cos\phi ) p_\phi/d $
where $d = \alpha_1\sin^2(\phi) + \alpha_2\cos^2(\phi)$.
The reduced Hamiltonian can be solved for $\eta_0$ on the section and thus
the integral $G$ can be written as a function of $(\phi, p_\phi)$ on the section.  Setting $g=\frac {2\alpha_2} {\alpha_2-\alpha_0} - \frac {\alpha_0} {\alpha_2(\alpha_2-\alpha_0)} j^2$, where we assume $h=1$, and rearranging the expression to find $p_\phi$ we have the section of the singular fibre is given by
\begin{equation}
p_\phi^2 = \left(\frac {2h\alpha_2} {\alpha_2-\alpha_0} - \frac {j^2} {(\alpha_2-\alpha_1)\sin^2(\phi)}\right)\frac {(\alpha_2-\alpha_0)(\alpha_2-\alpha_1)d} {\alpha_2(d-\alpha_0)} \cos^2(\phi).
\end{equation}

Figure~\ref{fig:fig2112a} shows the section for $j \ne 0$ for $\phi \ge 0$ which is the Fomenko atom $B$.  There is also a mirror image of $B$ if we plot the part of the graph for which $\phi \le 0$ .  However, for $j=0$ we get a different picture.  This is shown in figure~\ref{fig:fig2112b}.  Note that here we have the Fomenko atom $C_2$.  This is consistent because if we start with the graph for which $j \ne 0$, then as we decrease the value of $j$ the two copies of the atom $B$ approach each other, finally merging into one curve at $j=0$.  If we increase $j$ then as we approach the degenerate points the graphs of the $B$ atoms in the section reduce in size until all that is left of them at the degenerate points is two points in the $(\phi,p_\phi)$ plane on the $p_\phi=0$ axis.

Now the quotient with respect to $S$ has to be performed. In the new coordinates the $\Z_2$ action $S$ is $(\phi, p_\phi) \to (-\phi, -p_\phi)$. This action has no fixed points on the curve.  For $j=0$, under the action the $C_2$ is converted into $B$.  To see this, choose a fundamental region of the $\mathbb{R}^2$ plane, $\phi \ge 0$ say, then glue $p_\phi$ to $-p_\phi$ at $\phi=0$ and similarly at $\phi=\pi$.  For $j \ne 0$ the two $B$ merge into one so we again have $B$.

Since the reduced flow is transverse to the section on the singular fibre the complete reduced singular fibre is $B \times S^1$.  The singular fibre in full phase space is found by letting $\Phi$ act
on this set. There are no fixed points under $\Phi$ so every point will be multiplied by $S^1$.
Hence the singular fibre is $B \times T^2$.

For the degenerate points recall that the $B$ atoms reduced to two points which under the residual action map to a single point.  The singular fibre in reduced phase space is therefore $S^1$.  In full phase space we have a singular fibre of $T^2$.  This is logical because as we travel along the boundary curve, where the fibre is $T^2$, we reach the degenerate point and the fibre does not change.  This is expected behaviour for a degenerate point~\cite{bolsinov}.  When we enter the interior of the bifurcation diagram and the critical points become hyperbolic, the fibre changes to $B \times T^2$.  So we always have a $T^2$ in the fibre along the boundary curve.

\end{proof}

As in the case of the 121 ellipsoid we can define action variables for the 112 situation in terms of hyper-elliptic integrals.  However, we do not have an isolated singularity in this case nor do we have monodromy.  The fibre bundle over the regular points in each chamber of the bifurcation diagram will be trivial as in the generic case with distinct semi-axes.

\section{Ellipsoid with 211 symmetry}
Consider the geodesic flow on a three ellipsoid with the two smallest semi-axes equal, $0 < \alpha_0 = \alpha_1 < \alpha_2 < \alpha_3$, corresponding to the symmetry group action $SO(2) \times SO(1) \times SO(1)$. We call this ellipsoid the three dimensional prolate ellipsoid, in analogy to the two dimensional ellipsoid of revolution with a $SO(2) \times SO(1)$ symmetry.  The results are analogous to the 112 case, so we will only briefly summarise them. The integrals $F_0$ and $F_1$ from the generic case are not defined any more, but the singular terms cancel in the sum $G = F_0 + F_1$.  The other integrals $F_2$ and $F_3$ remain the same. 
The system is invariant under rotations in the $(x_0, x_1)$ plane and its cotangent lift, the $SO(2)$ group action being
\begin{equation}
\Phi(x, y; \theta) = (\tilde x, \tilde y)
\end{equation}
where
\begin{equation}
\begin{aligned}
\tilde x &= (x_0\cos \theta-x_1\sin \theta, x_0\sin \theta+x_1\cos \theta, x_2, x_3) \\
\tilde y &= (y_0\cos \theta-y_1\sin \theta, y_0\sin \theta+y_1\cos \theta, y_2, y_3) \, .
\end{aligned}
\end{equation}
As before the group action $\Phi$ is the flow generated by the angular momentum $J = x_0 y_1 - x_1 y_0$, which is once again a global action variable.

\begin{theorem} {\em Liouville Integrability} \label{thm:ALInt211}
The Geodesic flow on the ellipsoid with equal smallest semi-axes is Liouville integrable.  
Constants of motion are the energy $H=\frac {1} {2} \left(y_0^2+y_1^2+y_2^2+y_3^2\right)$, 
the angular momentum $J = x_0y_1 - x_1y_0$, 
and the third integral $G = F_0 + F_1$
\begin{equation}
G = y_0^2 + y_1^2 + \frac {\left(x_0y_2 - x_2y_0\right)^2} {\alpha_0 - \alpha_2} + \frac {\left(x_1y_2 - x_2y_1\right)^2} {\alpha_0 - \alpha_2} + \frac {\left(x_0y_3 - x_3y_0\right)^2} {\alpha_0 - \alpha_3} + \frac {\left(x_1y_3 - x_3y_1\right)^2} {\alpha_0 - \alpha_3}
\end{equation}
\end{theorem}
\begin{proof}
The proof is analogous to that of \eqref{thm:ALInt112}. 
\end{proof}
The group action $\Phi$ has the invariants
\begin{equation}
\pi_1 = x_0^2+x_1^2,\quad
\pi_2 = y_0^2+y_1^2,\quad
\pi_3 = x_0y_0 +x_1 y_1,\quad
\pi_4 = x_0 y_1 -x_1 y_0\,,
\end{equation}
related by $\pi_1\pi_2 -\pi_3^2 -\pi_4^2 =0$. The remaining variables
are trivial invariants. The fixed point of $\Phi$ is $x_0=x_1=y_0=y_1=0$.
As before, when $J=\pi_4 = j \not = 0$ this fixed point is not in $J^{-1}(j)$ and the reduction by
the $SO(2)$ symmetry leads to a smooth reduced system on $J^{-1}(j)/SO(2)$:

\begin{lemma}  \label{thm:Regredn211}
A set of reduced coordinates $(\xi_0, \xi_1, \xi_2, \eta_0, \eta_1, \eta_2)$ is defined on the reduced phase space $P_j = J^{-1}(j)/SO(2)$ by the formulae
\[
   \xi_0=\sqrt{\pi_1}, \quad \xi_1=x_2, \qquad \xi_2=x_3, \qquad
   \eta_0= \frac{\pi_3}{\sqrt{\pi_1}},\quad \eta_1=y_2, \qquad \eta_2 = y_3.
\]
The reduced coordinates satisfy the Dirac bracket in $ \R^6[\xi,\eta]$, i.e. $\{.,.\}_6$ as defined in \eqref{eqn:dirac}.
The mapping $R:\R^8[x,y] \to \R^6[\xi,\eta]$ is Poisson from $\R^8$ with $\{.,.\}_8$ to $\R^6$ with $\{.,.\}_6$ and the reduced system has reduced Hamiltonian
\[
\hat H = \frac12 ( \eta_0^2 +\eta_1^2+\eta_2^2)+ \frac{j^2}{2\xi_0^2} 
\]
and additional integral
\[
\hat G = \eta_0^2 + \frac{(\xi_0\eta_2 - \xi_2 \eta_0)^2}{\alpha_0 - \alpha_2} 
                            + \frac{(\xi_0\eta_3 - \xi_3 \eta_0)^2}{\alpha_0 - \alpha_3}
      + \frac{j^2}{\xi_0^2} \left( 1 +  \frac{\xi_2^2}{\alpha_0 - \alpha_2} 
                                                        + \frac{\xi_3^2}{\alpha_0 - \alpha_3} \right)  \,.
\]
\end{lemma}

\begin{proof}
This is analogous to the 112 situation.
\end{proof}

We get lemmas analogous to lemma \ref{thm:Separation112} for the Hamiltonian in local coordinates and lemma \ref{thm:SingRedn112} for the singular reduced space.  For the Liouville foliation we have:

\begin{figure}
\centerline{\includegraphics[width=8cm]{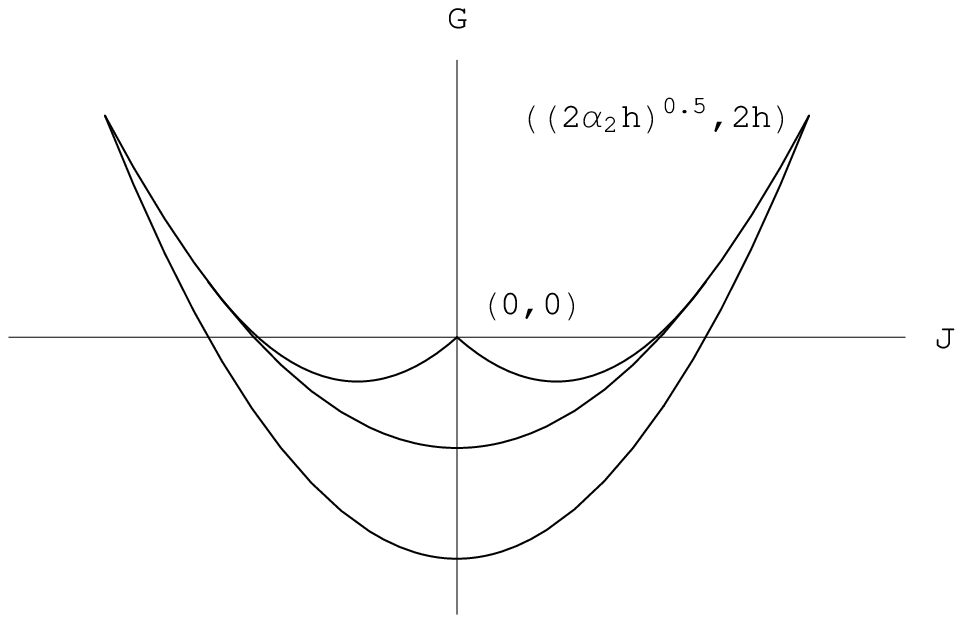}}
\caption{Bifurcation Diagram for the ellipsoid with $\alpha_0 = \alpha_1 < \alpha_2 < \alpha_3$ for $\alpha_0=1, \alpha_2=2, \alpha_3=3, h=1$.}
\label{fig:BiDi211}
\end{figure}

\begin{theorem} \label{thm:EM211}
The image of the energy momentum map ${\cal EM}$ for constant energy $H=h$ is the region in $\mathbb{R}^2$ bounded by the quadratic curves (see figure~\ref{fig:BiDi211})
\begin{equation}
g = \frac {2\alpha_0} {\alpha_0-\alpha_2} h + \frac {\alpha_2} {\alpha_0(\alpha_2-\alpha_0)} j^2,
\label{eqn:bdrycurves211A}
\end{equation}
\begin{equation}
g = \frac {2\alpha_0} {\alpha_0-\alpha_3} h + \frac {\alpha_3} {\alpha_0(\alpha_3-\alpha_0)} j^2,
\label{eqn:bdrycurves211B}
\end{equation}
\begin{equation}
g = -\sqrt{\frac {8\alpha_0h} {(\alpha_2-\alpha_0)(\alpha_3-\alpha_0)}} |j| + \frac {\alpha_2\alpha_3-\alpha_0^2} {\alpha_0(\alpha_2-\alpha_0)(\alpha_3-\alpha_0)} j^2.
\label{eqn:bdrycurves211C}
\end{equation}
Singular values of the energy momentum map are the boundary curves (elliptic), 
their transverse intersections (elliptic-elliptic), the boundary between the two chambers of the bifurcation diagram (hyperbolic), and the non-transverse intersections between the boundary curves (degenerate).
\end{theorem}

\begin{proof}
This is analogous to the 112 case.
\end{proof}
\noindent
Note that the curvature of the boundary curves in figure \ref{fig:BiDi211} is different to the curvature of the boundaries in figure \ref{fig:BiDi112} for the 112 ellipsoid because the boundary equations have different coefficients.  For the 211 ellipsoid, $G$ will always change sign due to the form of the quadratic equations for the boundaries, which are always negative for $J=0$ and always positive for the maximum $J$ value.  The tangency points can be positive or negative depending on the values of $\alpha_0, \alpha_2, \alpha_3$.

The fibres of the energy momentum map at the singular values in the bifurcation diagram are then classified by

\begin{theorem} \label{thm:SingFib211}
The singular fibres over the boundary curve \eqref{eqn:bdrycurves211A} of the image of the energy momentum map at constant energy, with the exception of its intersections, are two-dimensional tori $T^2$.  At each intersection point of the boundary curves \eqref{eqn:bdrycurves211A} and \eqref{eqn:bdrycurves211B} the singular fibre is $S^1$.  The singular fibre at the origin is two circles $S^1$.  For the boundary curves \eqref{eqn:bdrycurves211C} except at its intersections the singular fibre is two sets of two-dimensional tori $T^2$.  For the curve \eqref{eqn:bdrycurves211B} the singular fibre is a two-dimensional torus $T^2$ for each point on the boundary of the bifurcation diagram, $B \times T^2$ on the region which divides the bifurcation diagram into two chambers, and $T^2$ for the two degenerate points where the curve is tangential to \eqref{eqn:bdrycurves211C}, where $B$ is the Fomenko atom.  The multiplicity of the regular fibres $T^3$ is one for the chamber enclosed between \eqref{eqn:bdrycurves211A} and \eqref{eqn:bdrycurves211B} and two for the chamber enclosed between \eqref{eqn:bdrycurves211B} and \eqref{eqn:bdrycurves211C}.
\end{theorem}

\begin{proof}
This is analogous to the 112 case.
\end{proof}

\section{Ellipsoids with 13, 31 symmetry}

Consider the geodesic flow on the three dimensional ellipsoid corresponding to the group actions $SO(1) \times SO(3)$ and $SO(3) \times SO(1)$.  Taking $SO(1) \times SO(3)$ first we have
\begin{equation}
\frac {x_0^2} {\alpha_0} +  \frac {x_1^2} {\alpha_1} + \frac {x_2^2} {\alpha_1} + \frac {x_3^2} {\alpha_1} = 1
\end{equation}
where $\alpha_0 < \alpha_1$.  The Casimirs and the Dirac bracket are the same as before with $\alpha_1=\alpha_2=\alpha_3$.

For systems with non-commutative integrability, a more useful concept than the energy momentum map, when describing the dynamics of the system, is the energy Casimir map ${\cal EC}$.  This is defined here as ${\cal EC}: T^*\R^4 \rightarrow \R \times \R$, $(\mathbf{x},\mathbf{y}) \mapsto (H,J)$ where $H=\frac {1} {2} \left(y_0^2+y_1^2+y_2^2+y_3^2\right)$ is the energy and $J$ the total angular momentum arising from the symmetry group action of $SO(3)$.
\begin{theorem} {\em Non-commutative Integrability.} \label{thm:neh}
The equations of the geodesic flow on the ellipsoid with a symmetry corresponding to $SO(1) \times SO(3)$ are non-commutative integrable; namely two independent involutive integrals are the energy $H$ and total angular momentum $J$, and two independent but non-involutive integrals given by angular momenta $L_{12}, L_{13}$.  The fibre over a regular point of the energy Casimir map $(H,J)=(h,j)$ is a fibre bundle with base space $S^2$ and fibre $T^2$.
\end{theorem}
\begin{proof}
The $SO(3)$ group action gives rise to three angular momenta $L_{12}, L_{13}, L_{23}$ (defined by $L_{12}=x_1y_2-x_2y_1$ etc.) belonging to the Lie algebra $so(3)^*$.  These are all functionally independent.  The total angular momentum $J$ is defined by
\begin{equation}
J^2 = L_{12}^2 + L_{13}^2 + L_{23}^2.
\end{equation}
Taking $J$ as an integral leaves two functionally independent $L_{ij}$.  By a direct calculation
\begin{equation}
\{H, J\} = 0, \qquad \{H, L_{ij}\} = 0, \qquad \{J, L_{ij}\} = 0, \qquad \{L_{12}, L_{13}\} = x_2y_3 - x_3y_2.
\end{equation}
Hence we have two independent involutive integrals and two independent but non-involutive integrals.  We apply Nekhoroshev's theorem~\cite{neh72} on non-commutative integrability to show that motion takes place on invariant two tori $T^2$.  However, this is when we fix the constants of motion.  Fixing $J$, $L_{12}$ and $L_{13}$ is equivalent to fixing $L_{12}$, $L_{13}$ and $L_{23}$, assuming that they are in the correct range (i.e. obviously $L_{23}$ is defined by $L_{23}^2=J^2-L_{12}^2-L_{13}^2$ and this must be positive).  Now the individual angular momenta $L_{ij}$ belong to $so(3)^*$ which is isomorphic to $\mathbb{R}^3$ with the usual isomorphism of multiplication being mapped to the vector product.  But this allows the $L_{ij}$ to vary over $\mathbb{R}^3$ so is not quite the situation here.  For the energy Casimir map where we fix $(H,J)=(h,j)$, the total angular momentum $J$ is fixed so the $L_{ij}$ are free to vary but are constrained on a sphere $S^2$.  So each regular fibre of the energy Casimir map is a fibre bundle with fibre $T^2$ and base space $S^2$.
\end{proof}
An analogous situation to this result was described by Fasso~\cite{fasso02} for broadly integrable Hamiltonian systems, of a meadow of actions in which there are flowers whose petals are tori parameterised by angles conjugate to the actions and whose centres are coadjoint orbits.

For the group action $SO(3)$ we have the invariants
\begin{equation}
\pi_1 = x_1^2 + x_2^2 + x_3^2,
\end{equation}
\begin {equation}
\pi_2 = y_2^2 + y_3^2 + y_4^2,
\end{equation}
\begin{equation}
\pi_3 = x_1y_1 + x_2y_2 + x_3y_3.
\end{equation}
The variables $x_0, y_0$ are trivial invariants of the group action.  We use the invariants to define variables in a reduced system as follows:
\begin{lemma}  \label{thm:Regredn13}
A set of reduced coordinates $(\xi_0, \xi_1, \eta_0, \eta_1)$ is defined on the reduced phase space $P_j=J^{-1}(j)/SO(3)$ by the formulae
\begin{equation}
   \xi_0=x_0, \quad \xi_1=\sqrt{\pi_1}, \qquad
   \eta_0=y_0,\quad \eta_1 =  \frac{\pi_3}{\sqrt{\pi_1}}
   \nonumber
\end{equation}
The reduced coordinates satisfy the Dirac bracket in $\R^4[\xi,\eta]$, i.e. $\{.,.\}_4$ as defined in \eqref{eqn:dirac}.
The mapping $R:\R^8[x,y] \to \R^4[\xi,\eta]$ is Poisson from $\R^8$ with $\{.,.\}_8$ to $\R^4$ with $\{.,.\}_4$ and the reduced system has reduced Hamiltonian
\begin{equation}
\hat H = \frac12 ( \eta_0^2 +\eta_1^2)+ \frac{j^2}{2\xi_1^2}
\nonumber
\end{equation}
representing a one degree of freedom system on an ellipse.
\end{lemma}
\begin{proof}
The proof is analogous to lemma \ref{thm:Regredn112}.
\end{proof}
We can now carry out singular reduction by
\begin{lemma}
The reduced phase space $P_j=J^{-1}(j)/SO(3)$ is an open subset of $T^*S^1$, diffeomorphic to $\R^2$, in the case that $j \ne 0$. For $j=0$ the reduced space is two dimensional with a conical singularity.
\end{lemma}
\begin{proof}
The Casimirs and the Hamiltonian are linear in the invariants.  The reduced phase space is a subset of $\mathbb{R}^5[x_0,y_0,\pi_1,\pi_2,\pi_3]$ and defined by the Casimirs, the relation $\pi_1\pi_2 - \pi_3^2 = j^2$ and the inequalities $\pi_1 \ge 0$, $\pi_2 \ge 0$.  Eliminating the invariants $\pi_1$ and $\pi_3$ using the Casimirs gives
\begin{equation}
\alpha_1\left(1-\frac {x_0^2} {\alpha_0}\right)\pi_2 - \frac {\alpha_1^2x_0^2y_0^2} {\alpha_0} = j^2,
\end{equation}
which defines the reduced phase space $P_j = J^{-1}(j)/SO(3)$.  As before this is smooth except for $j=0$.  For $j \ne 0$ we have $\pi_2>0$ and the reduced phase space is one sheet of a two sheeted hyperboloid, which is diffeomorphic to $\mathbb{R}^2$.  We can also see this by looking at the two Casimirs in the reduced variables $(\xi_0, \xi_1, \eta_0, \eta_1)$ which give an open subset of $T^*S^1$.  We only have an open subset and not the whole cotangent bundle as $\xi_1>0$.  For $j=0$ we have $\pi_2 \ge 0$ and here we get a conical singularity in the reduced phase space.
\end{proof}

For the singular fibres:
\begin{theorem} {\em Singular Fibres}
For $J=0$ we have a singular fibre $S^2 \times S^1$ and for $J^2 = 2\alpha_1h$ we have a singular fibre $SO(3)$.
\end{theorem}
\begin{proof}
For the geodesic flow on the ellipsoid corresponding to a symmetry group $SO(1) \times SO(3)$ we have $0 \le J^2 \le 2\alpha_1h$.  Eliminating the invariants in the relation using the the Casimirs and the Hamiltonian gives
\begin{equation}
\alpha_1\left(1 - \frac {x_0^2} {\alpha_0} \right) \left(2h - y_0^2 \right) - \frac {\alpha_1^2x_0^2y_0^2} {\alpha_0^2} = J^2
\label{eqn:so3reln}
\end{equation}
Differentiating this relation with respect to the variables $x_0$ and $y_0$ and setting the derivatives equal to zero gives a critical point at $x_0=0, y_0=0$.  Here the invariants are $\pi_1 = \alpha_1$, $\pi_2 = 2h$ and $\pi_3 = 0$.  This gives us the singular value $J^2 = 2\alpha_1h$.  Inserting this value into (\ref{eqn:so3reln}) gives
\begin{equation}
y_0^2\left(1 + \frac {(\alpha_1 - \alpha_0)} {\alpha_0^2} x_0^2\right) = - \frac {2hx_0^2} {\alpha_0} \le 0
\end{equation}
and hence the only solution is $x_0=0, y_0=0$.  Thus the singular fibre is defined by the equations
\begin{equation}
x_1^2 + x_2^2 + x_3^2 = \alpha_1
\end{equation}
\begin{equation}
y_1^2 + y_2^2 + y_3^2 = 2h
\end{equation}
\begin{equation}
x_1y_1 + x_2y_2 + x_3y_3 = 0.
\end{equation}
But this is just the unit circle bundle over $S^2$ which is known to be $SO(3)$.
\\
\\
Note that the relation we differentiated to find the singular points was equal to $J^2$.  This means that another singular point arises due to the problem with differentiability at $J=0$.  Setting (\ref{eqn:so3reln}) equal to zero and rearranging gives
\begin{equation}
y_0^2 = \frac {\frac {2h\alpha_0^2} {\alpha_1 - \alpha_0} - \frac {2h\alpha_0} {\alpha_1 - \alpha_0} x_0^2} {x_0^2 + \frac {\alpha_0^2} {\alpha_1 - \alpha_0} }.
\end{equation}
The curve represented by this equation is homoemorphic to a circle $S^1$ in reduced space.  For any point on this circle we choose any point in the inverse image of the reduction map and look at the orbit of $SO(3)$.  As the point is never a fixed point of the group action, or we would require $(x_1, x_2, x_3) = (0, 0, 0)$ and $(y_1, y_2, y_3) = (0, 0, 0)$ which is not possible, the orbit is always $S^2$ and hence the singular fibre is $S^2 \times S^1$.
\end{proof}
\noindent
For the $SO(3) \times SO(1)$ group action, the situation is similar to the previous case.  We have

\begin{theorem} {\em $SO(3) \times SO(1)$ Ellipsoid.}
For the geodesic flow on the ellipsoid corresponding to a $SO(3) \times SO(1)$ symmetry, the behaviour of the flow is analogous to that of the $SO(1) \times SO(3)$ ellipsoid.  For total angular momentum $J$ we have $0 \le J^2 \le 2\alpha_0h$.  For $J=0$ we once again have a singular fibre $S^2 \times S^1$, for $J^2 = 2\alpha_0h$ we have a singular fibre $SO(3)$, and for intermediate values of $J^2$ the points are regular and the regular fibres are a $T^2$ bundle over $S^2$.
\end{theorem}

The ellipsoid with a $SO(4)$ symmetry is just the geodesic flow on $S^3$, which is well known.  It has been described in detail by Cushman~\cite{cushman97}.

\bibliographystyle{plain}
\bibliography{Paperbibfile}

\end{document}